\documentclass{emulateapj}
\usepackage{amsmath}

\def\head{ \vbox to 0pt{\vss \hbox to 0pt{\hskip 440pt\rm
      LA-UR-08-07210\hss} \vskip 25pt}}

\begin{document}

\head

\shorttitle{Fossil Systems in the 400d Cluster Catalog}
\shortauthors{Voevodkin et al.}

\title{Fossil Systems in the 400d Cluster Catalog}

\author{
  Alexey~Voevodkin\altaffilmark{1,2}, Konstantin~Borozdin\altaffilmark{1},
  Katrin~Heitmann\altaffilmark{1}, Salman~Habib\altaffilmark{1},
  Alexey~Vikhlinin\altaffilmark{3,2},
  Alexander~Mescheryakov\altaffilmark{2,3}, Allan~Hornstrup\altaffilmark{4}
}
\altaffiltext{1}{
  Los Alamos National Laboratory, Los Alamos, NM, 87545
}
\altaffiltext{2}{
  Space Research Institute (IKI), Profsoyuznaya 84/32, Moscow, Russia
}
\altaffiltext{3}{
  Harvard-Smithsonian Center for Astrophysics, 60 Garden St.,
  Cambridge, MA 02138
}
\altaffiltext{4}{
  Technical University of Denmark, National Space Institute, Juliane
  Maries Vej 30, 2100 Copenhagen 0, Denmark
}
\begin{abstract}
  We report the discovery of seven new fossil systems in the 400d
  cluster survey. Our search targets nearby, $z\le0.2$, and X-ray
  bright, $L_X\ge 10^{43}$ erg~sec$^{-1}$, clusters of galaxies.
  Where available, we measure the optical luminosities from Sloan
  Digital Sky Survey images, thereby obtaining uniform sets of both
  X-ray and optical data. Our selection criteria identify 12 fossil
  systems, out of which five are known from previous studies.  While
  in general agreement with earlier results, our larger sample size
  allows us to put tighter constraints on the number density of fossil
  clusters.  It has been previously reported that fossil groups are
  more X-ray bright than other X-ray groups of galaxies for the same
  optical luminosity.  We find, however, that the X-ray brightness of
  massive fossil systems is consistent with that of the general
  population of galaxy clusters and follows the same $L_X-L_{\rm opt}$
  scaling relation.
\end{abstract}


\section{Introduction}

Studies of clusters and groups of galaxies have revealed an
interesting class of objects -- relaxed, X-ray bright systems,
dominated by a giant elliptical galaxy at the center. These systems
are usually called ``fossil groups''~\citep{1994Natur.369..462P}, or,
if massive enough, ``fossil clusters''
~\citep{2006AJ....132..514C}. The interesting properties of these
systems are attributed to their dynamical history, including a lack of
galaxy mergers for a long time prior to observation. Because of their
quiescent state, fossil groups and clusters (hereafter FGs) are
``frozen in time'' as compared to other systems.  In X-ray
observations FGs appear relaxed, without any obvious sign of recent
merging processes.

The number of known FGs is very small, and only a handful of them have
been studied in some detail.  Known FGs are massive systems with
masses in the range typical for rich groups or for poor clusters of
galaxies; they share several scaling relations with other groups and
clusters~\citep{2007MNRAS.377..595K}. Their central galaxies have
properties very similar to those of the brightest cluster
galaxies. The merger histories underlying the origin of the central
galaxies in FGs continues to be discussed in the literature (see
e.g.~\citealt{2003MNRAS.343..627J} and~\citealt{1999ApJ...514..133M} for
different points of view).

The definition of fossil groups and clusters remains somewhat
ambiguous, as different researchers use somewhat different criteria to
identify these systems. Additionally, systems of a possibly ``fossil''
nature have been called overluminous elliptical galaxies (OLEGs,
~\citealt{1999ApJ...520L...1V}), or isolated overluminous elliptical
galaxies (IOLEGs, ~\citealt{2004AdSpR..34.2525Y}).  Even if the
nomenclature differs, all these objects are in essence gravitationally
bound groups or clusters of galaxies, with the central elliptical
galaxy dominating all others galaxies in the system. 

In this work we
loosely follow the definition of FGs as given
in~\cite{2003MNRAS.343..627J} (hereafter J03), viz., a fossil group is
an extended X-ray bright object ($L_{X,\rm
  Bol}\ge0.25\times10^{42}\,h^{-2}$ erg~s$^{-1}$), with $\Delta
m_{12}\ge2$ for galaxies lying inside half of the virial radius, where
$\Delta m_{12}$ is the absolute magnitude difference in the R-filter
between the first and second brightest galaxy. This definition is
phenomenological, and it was applied by J03 to identify five such
objects.  One obvious deficiency of this definition is that it imposes
an artificially sharp $\Delta m_{12}$ threshold (representing the tail
of the Schechter function), whereas no such sharp boundary can be
found in the $\Delta m_{12}$ distribution in clusters
\citep{2006ApJ...637L...9M, 2008arXiv0812.2929L}.  Therefore, for any
chosen threshold of $\Delta m_{12}$ there will always be systems very
close to the boundary, and their identification would depend on the
accuracy of photometric measurements in optical observations.  There
is a very similar, but perhaps even more difficult, problem associated
with the measurement of the virial radius.  In practice, some
researchers prefer instead to consider galaxies within some fixed
radius \citep{2007AJ....134.1551S, 2008arXiv0812.2929L}.  Finally,
group membership cannot be reliably determined for individual
galaxies.  Still, the J03 criteria have the advantage of being
definitive, and remain the the most popular choice for FG
identification in observations.  We re-evaluate these criteria as part
of our analysis.

In contrast to considering fossil systems as a separate category some
researchers speculate that there may be a ``fossil phase'' in the life
of many clusters, with an absence of significant mergers for a long
time, enough for cluster relaxation~\citep{2008MNRAS.386.2345V}.  A
bright galaxy may consequently fly into the inner cluster part, and
the object will now appear to the observer as a normal group or
cluster of galaxies.

It is clear that to gain a better understanding of the nature of FGs
more of these systems have to be observed, and a comparison of FGs
with normal groups or clusters of galaxies should be carried out in a
uniform manner, free from selection effects as much as possible. We
aim here to identify new FGs and to compare their properties with
those of other systems in the same mass range.

For our search we use the 400d X-ray survey of galaxy
clusters~\citep{2007ApJS..172..561B}.  The advantage in using X-ray
surveys is that in this case our selection is based on the presence of
an extended X-ray source, indicating that the object is a
gravitationally bound system, and not a chance
superposition of galaxies. We can then proceed to identify FGs in the
optical band.  The 400d survey provides an excellent database for
this work, since almost every cluster has a CCD image in the R-band.
We use different optical data to identify FGs in the 400d survey, but
our comparison of FGs with other groups and clusters has been mostly
enabled by the Sloan Digital Sky Survey (hereafter SDSS) DR6,
providing us with a uniform data set both in the X-ray and in the
optical. The survey area as a function of flux is well calibrated,
allowing us to put constraints on the FG number density. (A subcatalog
of the 400d survey -- the 160d survey ~\citep{1998ApJ...502..558V} was
previously used to identify OLEGs.)

The paper is organized as follows. In Section~\ref{sec2} we describe
the available data and the methodology underlying our FG search. In
Sections~\ref{sec3}-\ref{sec5} we present our results: seven new
FGs, constraints on the number density of FGs, and an $L_X-L_r$
correlation. We summarize our findings in Section~\ref{sec6}. The
Appendix contains a detailed description of all individual fossil
systems we study in this paper.

In this work we consider luminous systems with X-ray luminosities
$L_X\ge 10^{43}$ erg~sec$^{-1}$. Masses of these objects, estimated
from their X-ray luminosities, correspond to rich groups or poor
clusters of galaxies.  We generally refer to these systems as clusters
to highlight the fact that we study here the bright and massive end of
the distribution of fossil systems.  We plan to study less massive
systems in future work.

Where we need to assume a cosmology, we choose a $\Lambda$CDM model,
with $\Omega_M = 0.3$, $\Omega_\Lambda = 0.7$ and $h=0.71$, the Hubble 
constant, measured in units of
100~km~s$^{-1}$~Mpc$^{-1}$. X-ray luminosities and fluxes are given in
the 0.5--2.0 keV band. We use r-filter apparent SDSS magnitudes, and
we employ galactic extinction corrections and transform them to AB
systems to compute galaxy luminosities. The difference of the apparent
magnitudes in the R-filter is used in other cases.

\section{Data analysis}\label{sec2}

Our sample is based on the 400d cluster survey
catalog~\citep{2007ApJS..172..561B}. This is a catalog of X-ray
selected clusters of galaxies from \emph{ROSAT} PSPC high latitude
pointing observations. The area of the sky covered by the survey is
397 square degrees. It includes clusters with fluxes higher than
$1.4\times10^{-13}$ erg~s$^{-1}$~cm$^{-2}$ -- 242 clusters in total.
Approximately a third of the clusters have redshifts higher than
$z=0.3$.  Every object has a measured redshift, X-ray flux, and X-ray
luminosity. Most of the objects have CCD R-band images mainly obtained
with the Russian-Turkish 1.5-m telescope in the North and the Danish
1.54-m telescope in the South\footnote{Some observations were also
  performed at the Multiple Mirror Telescope, at the ESO 3.6m, and at
  the FLWO 1.2-m telescope. Detailed information can be found online:\\
  http://hea-www.harvard.edu/400d/catalog/table\_cat.html}.

The 160d catalog~\citep{1998ApJ...502..558V} -- a subcatalog of the
400d catalog -- was previously used for the search of FGs
~\citep{1999ApJ...520L...1V}. In this work we have decided to use the
same selection criteria, i.e., we select clusters with $z\le0.2$ and
$L_X\ge10^{43}$~erg~s$^{-1}$. These cuts provide us with clusters of
the same richness (or higher) as poor Abell clusters, being visible on
CCD images as galaxy concentrations (see~\citealt{1999ApJ...520L...1V}).
Applying these cuts to the complete 400d catalog yields 75 candidates
for our FG search.

In the next step we visually inspect the optical images of the 75
candidates. It turns out that many of them do not have optical images
of sufficient quality for our purposes. Either their field of view is
not large enough to cover the desired circle with radius $0.5r_{\rm
  vir}$ or the quality of the CCD image does not allow for reliable
photometry, or it was originally just a scanned photographic plate
(DSS2). We use archival data from the SDSS
DR6~\citep{2008ApJS..175..297A}, whenever available. We have SDSS data
for 38 clusters from the original sample; for the remaining clusters,
we have to work with lower quality optical information.

\subsection{Identification of Fossil Groups}

\begin{figure*}
  \plottwo{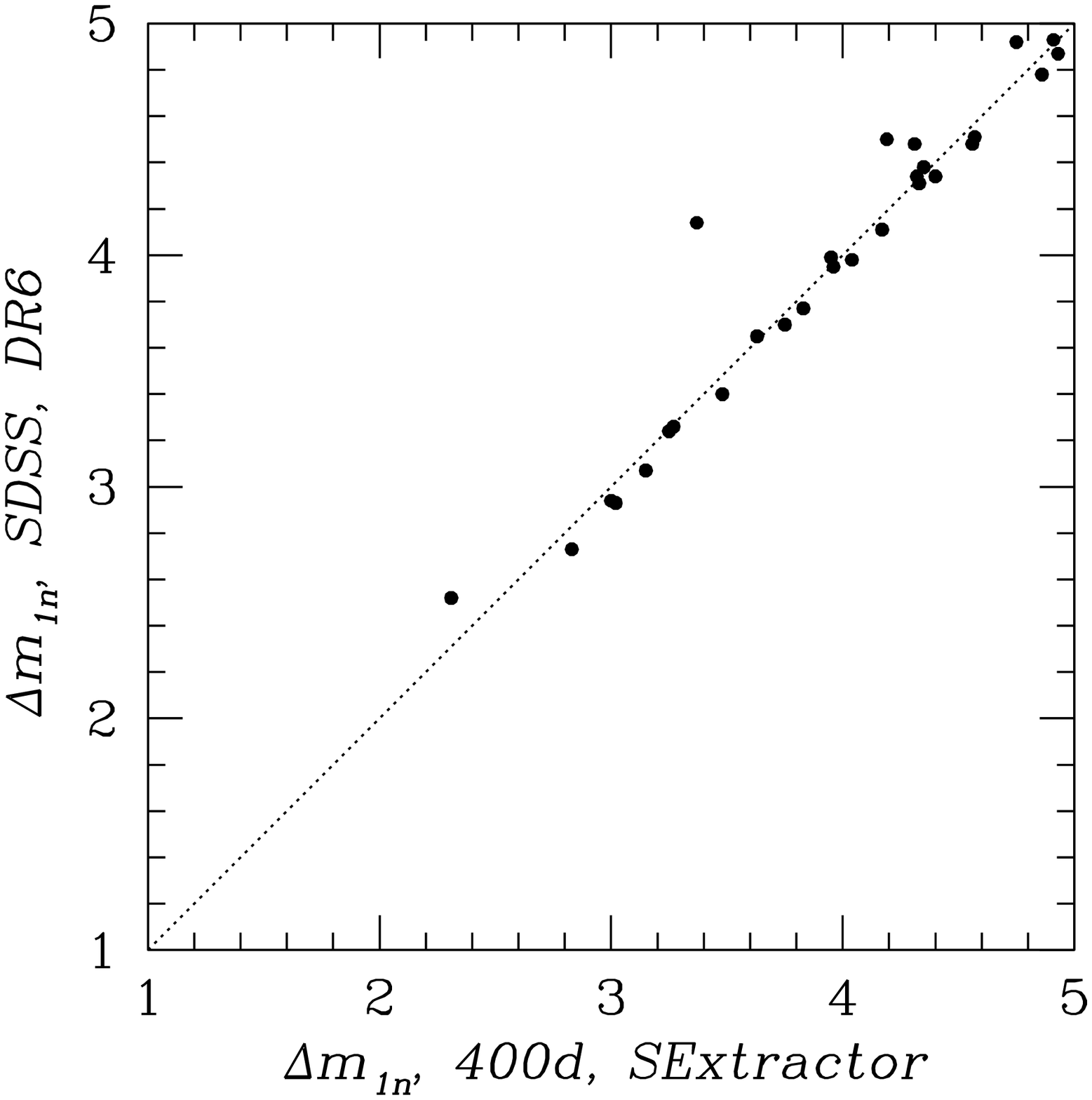}{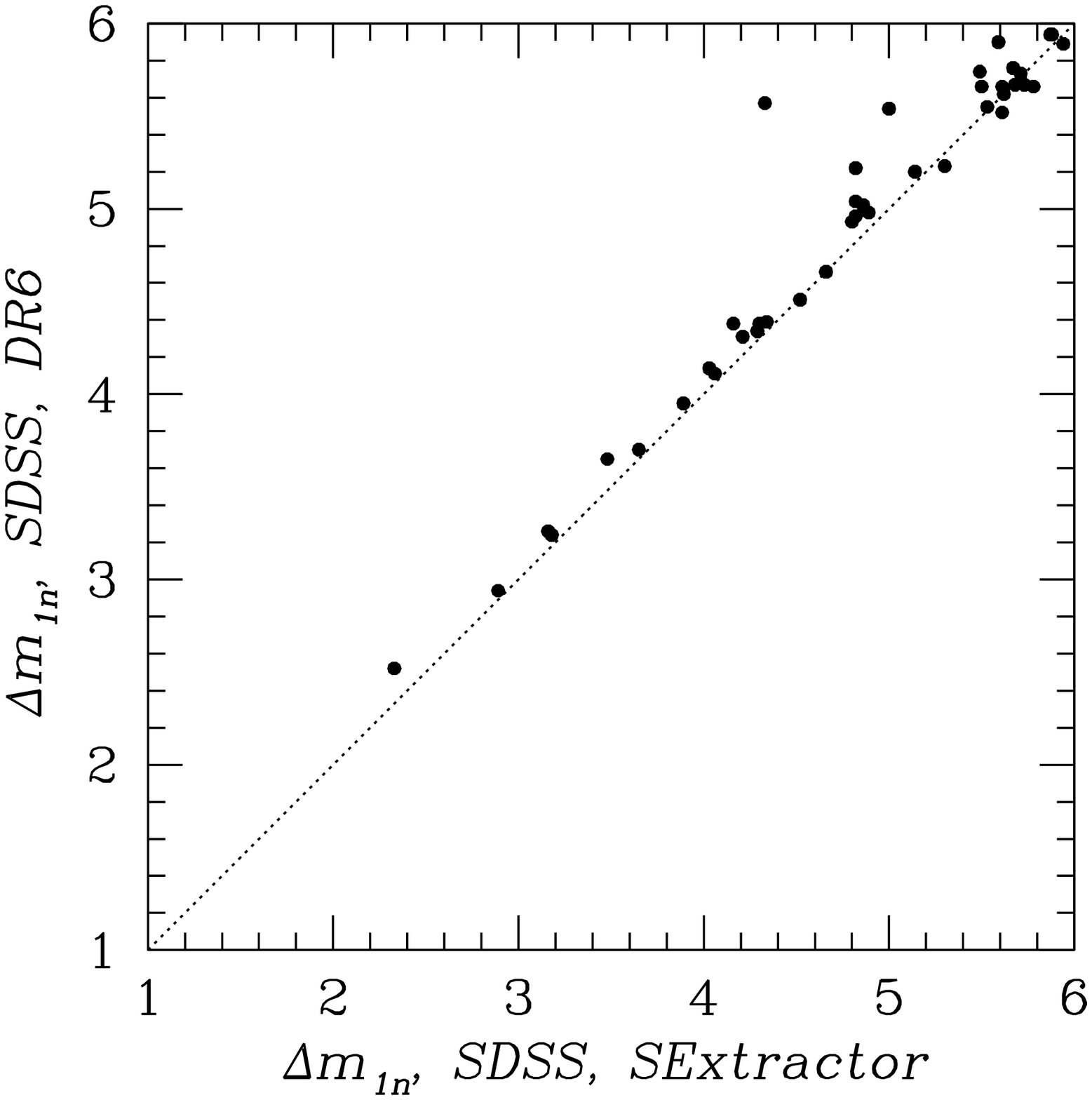}
  \caption{Comparison of the r-filter photometry from SDSS DR6 with
    other photometric measurements for the cluster cl1159p5531.  Dots
    represent the difference in apparent magnitude between the central
    galaxy of the cluster and other galaxies.  \emph{Left panel}:
    Comparison of the r-filter photometry from SDSS DR6 with
    photometry done by SExtractor on the R-band image from the FLWO
    telescope.  \emph{Right panel}: Photometry from the SDSS DR6
    compared with the photometry done by SExtractor on the corrected
    frame r-filter SDSS image. Note that due to different areas of
    coverage, not all galaxies are the same for the two panels.}
 \label{fig:photom}
\end{figure*}

In 2003, nine years after the discovery of the first fossil system
by~\cite{1994Natur.369..462P} and several other detections of fossil
systems, J03 proposed the first concrete definition of FGs. Since
then, this definition has been adopted by most of the community, and
new FGs have been detected by various research groups. A major
difficulty from the observational perspective is to ensure that all
galaxies belonging to the system have been identified. For example,
RX~J$1552.2+2013$ was identified (J03) as an FG and later studied
again~\citep{2006AJ....131..158M} in detail. It was only recently
found~\citep{2008arXiv0809.2036Z} that this system does not in fact
obey the Jones criteria: a bright galaxy within half of the virial
radius had not been seen in previous exposures, which only covered a
fraction of the half-virial radius circle around the central
galaxy. As another example, the object RX~J$1159.8+5531$ was studied
by~\cite{1999ApJ...520L...1V} and identified as an OLEG. However, it
was found from SDSS data~\citep{2008arXiv0809.3483D} that a bright
galaxy inside half of the virial radius had been missed previously,
and therefore the system should not have qualified as an FG. We
discuss these examples to caution the reader that even if a system has
been previously identified as an FG, new studies with better data may
reveal that the identification is erroneous.

Another question concerns the accuracy of the photometry obtained for
the galaxy members of the system. The standard definition of an FG
requires $\Delta m_{12}\ge2$, but depending on the observational
conditions and the methods used, the galaxy flux measurements for the
same object might differ. We therefore view the optical criteria
suggested by J03 -- $\Delta m_{12}\ge2$ within half the virial radius
-- as guidelines rather than strict requirements.

As a first check for our sample of 75 FG candidates, we consider the
systems previously identified as FGs.  Such systems allow us to
cross-check our observational results with the analysis carried out by
other groups and to re-evaluate the Jones criteria. We find six known
FGs present in our sample (see Table~4 in~\citealt{2006AJ....131..158M}).
They are: cl1159p5531, cl1340p4017, cl1416p2315, cl1552p2013,
cl2114m6800, and cl2247p0337\footnote{Here and below we use object
  names from the 400d catalog. The meaning of these names is
  straightforward: the first four digits represent the right
  ascension, the last four, the declination, p and m denote plus and
  minus signs for the Northern and Southern hemispheres,
  respectively.}.  The first four of these systems have been observed
by the SDSS and therefore good quality optical data is available.

It is known that SDSS magnitudes of bright galaxies may be
underestimated due to an overestimation of the background
level~\citep{2008ApJS..175..297A}. Therefore, we double-check the
photometry for one of the known FGs -- cl1159p5531.  In
Figure~\ref{fig:photom} we compare the difference in apparent
magnitudes between the brightest and all other galaxies, $\Delta
m_{1n}$, lying inside $r_{500}$ for two different cases. In the left
panel we compare the photometry obtained with
SExtractor~\citep{1996A&AS..117..393B} on the R-filter image measured
by the FLWO 1.2m telescope vs. SDSS DR6 r-filter photometry. In the
right panel we compare photometry obtained with SExtractor on the
corrected frame SDSS r-filter image (run 2821, frame 184) again
vs. SDSS DR6 r-filter photometry. This figure demonstrates that for
this group we do not need to correct the SDSS photometry. We also
confirm that the photometry obtained from SExtractor on SDSS corrected
frame images for several other groups also does not need a
correction. Therefore, everywhere below we do not apply any
corrections to the SDSS data. However, we will establish criteria for
our FG search that take into account possible inaccuracies in SDSS
photometry by which $\Delta m_{12}$ might be underestimated.

In order to identify a system as an FG, we first have to
establish a search radius to identify group members and measure
$\Delta m_{1n}$. Following the standard definition, the search radius
should be some fraction of the virial radius. Using the cluster X-ray
luminosities from the 400d catalog we can easily estimate $r_{500}$
for every cluster in our sample from~\citep{2008arXiv0805.2207V}:
\begin{equation}\label{r500}
  r_{500} = 8.17\times10^{-7}\left(\frac{L_X}{{\rm erg\ 
        s}^{-1}} \right)^{0.21} \left(\frac{h}{0.72}\right)^{-0.59}
  E(z)^{-1.05},
\end{equation}
where $r_{500}$, measured in kpc, is the radius inside which the mean
cluster density is 500 times higher than the critical density of the
Universe. $r_{500}$ can be determined from Eq.~(\ref{r500}) with an
uncertainty of approximately 8\%.  The relation between $r_{500}$ and
the virial radius is given by $r_{500}\approx 0.6r_{\rm vir}$.
Therefore, $r_{500}$ covers a slightly larger area than suggested by
the Jones criteria. We decide to keep this more conservative limit for
our search radius as a starting point.

\begin{figure*}
  \centerline{
    \includegraphics[width=0.25\linewidth]{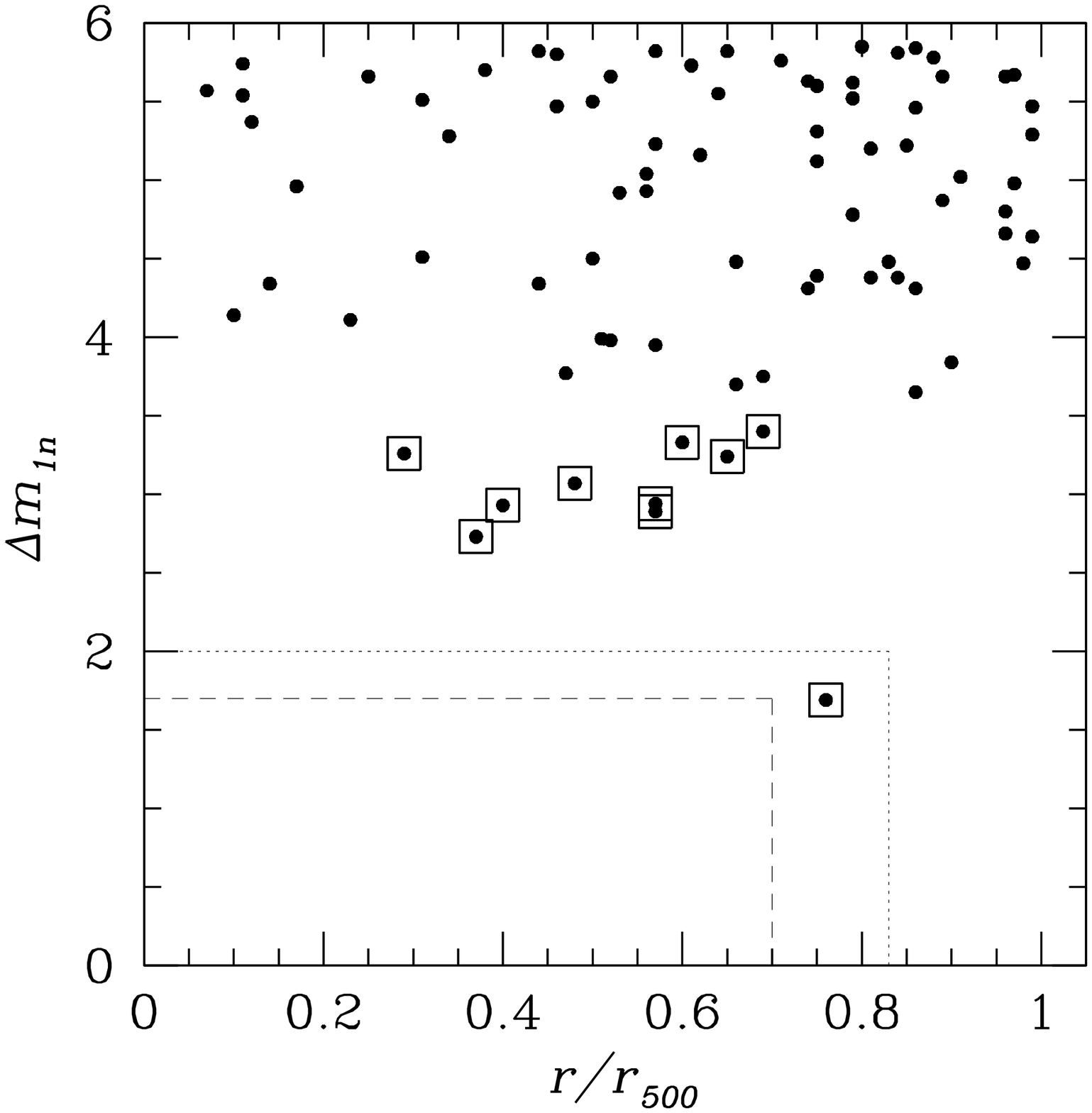}
    \includegraphics[width=0.25\linewidth]{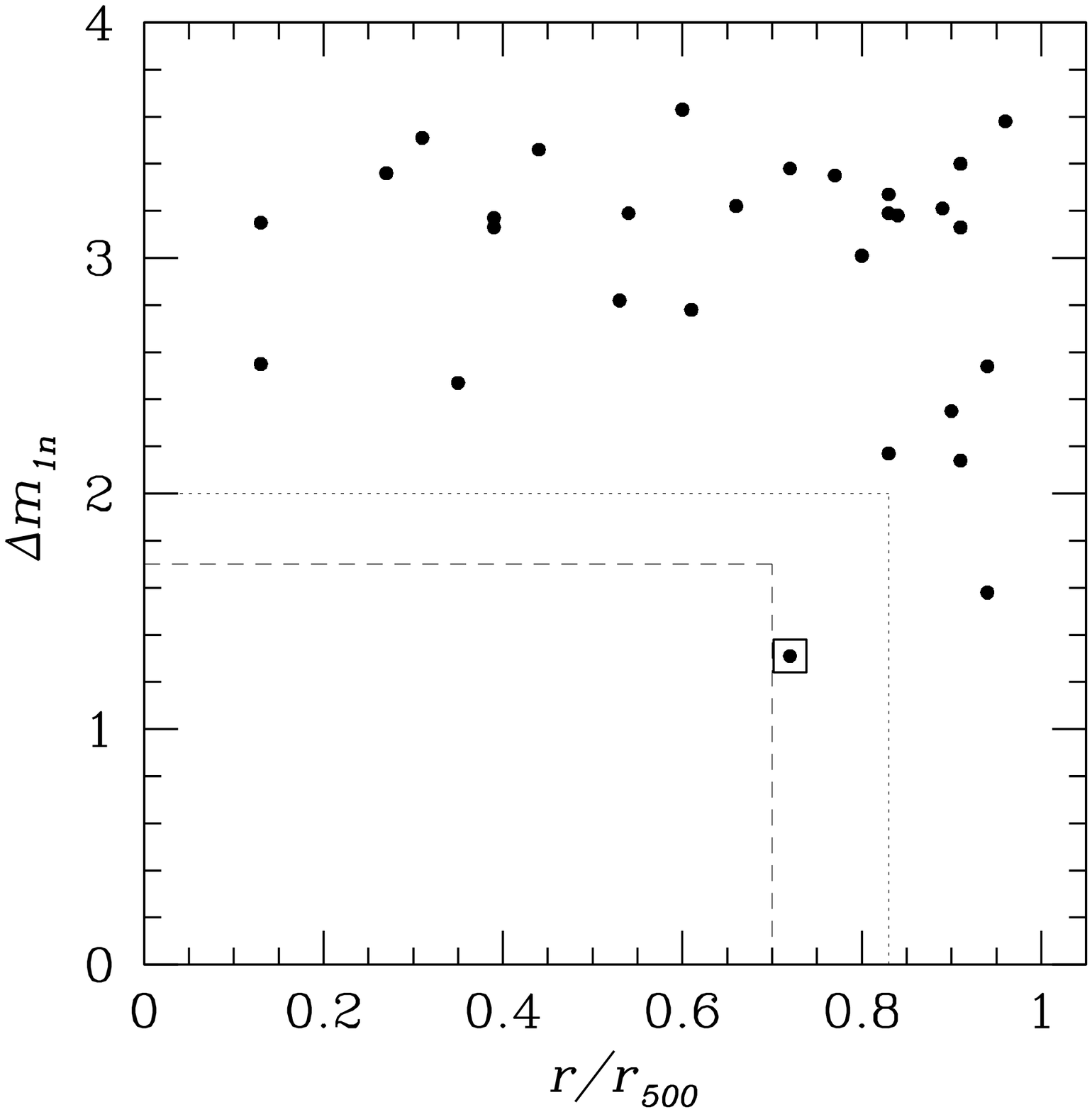}
    \includegraphics[width=0.25\linewidth]{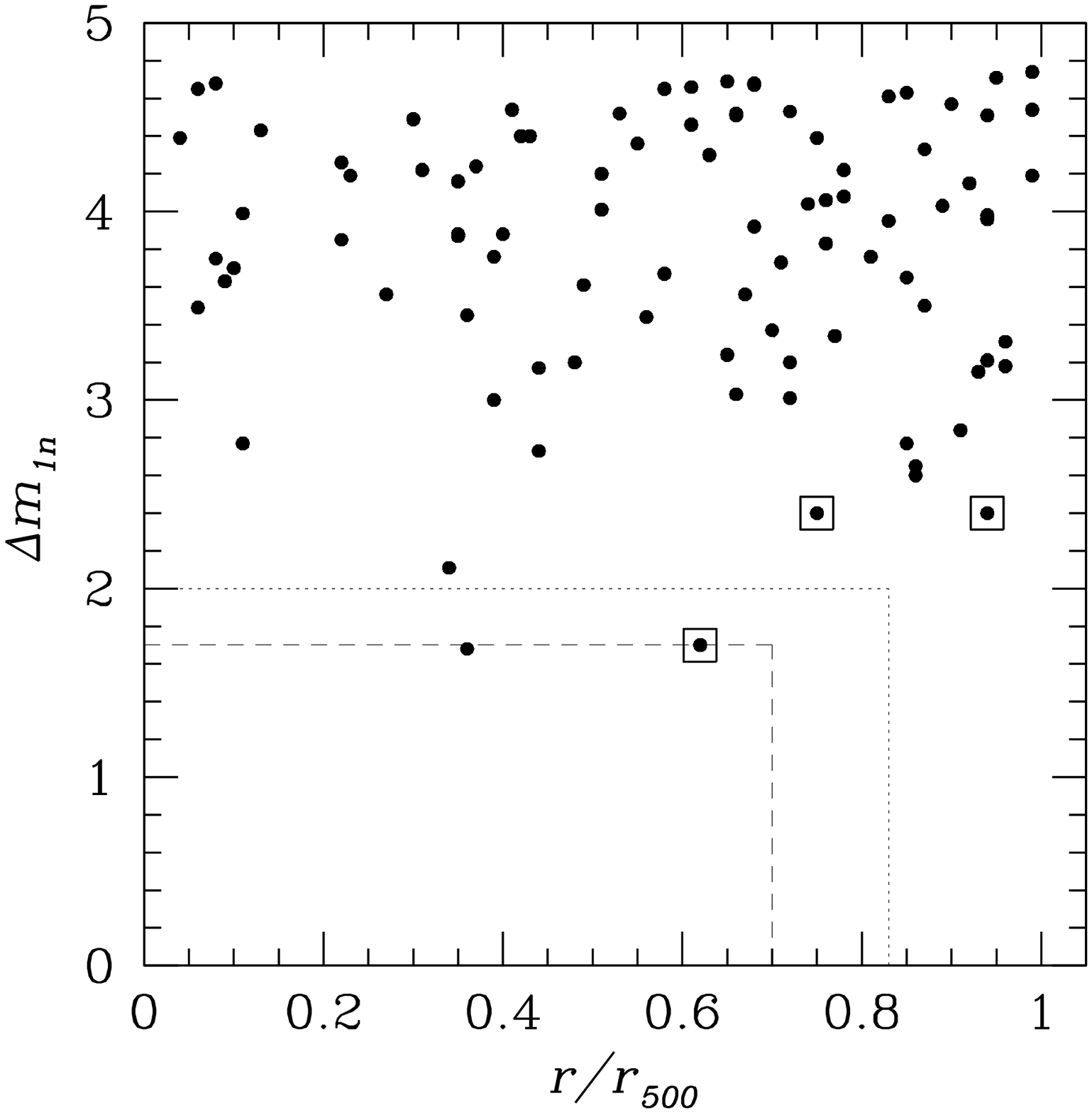}
    \includegraphics[width=0.25\linewidth]{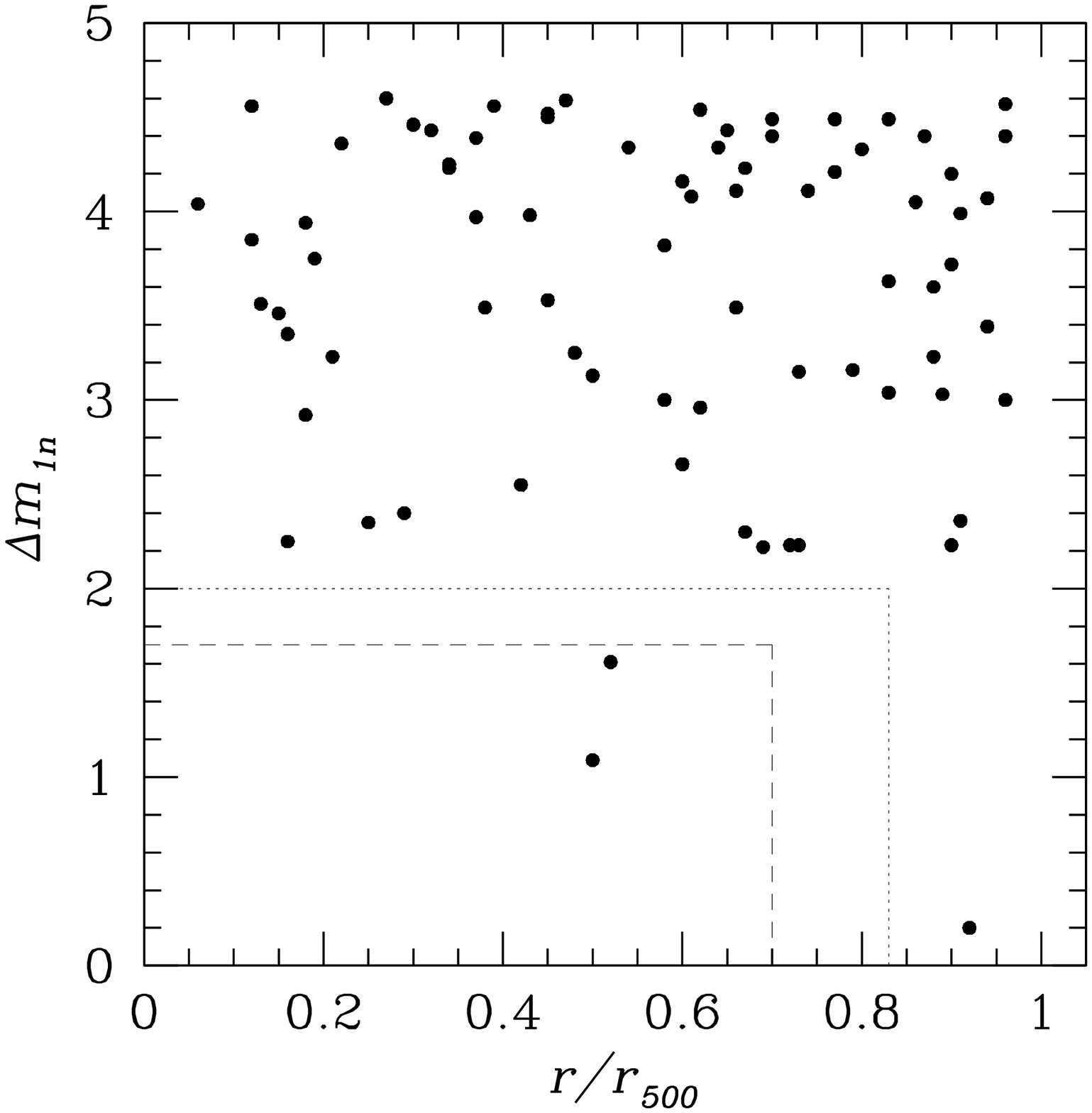}
  }
  \caption{Difference in magnitude $\Delta m_{1n}$ between the central
    galaxy of the cluster and other individual galaxies, as a function
    of distance to the X-ray center. The four panels correspond to
    clusters (from left to right) cl1159p5531, cl1340p4017,
    cl1416p2315, and cl1552p2013, identified earlier as fossil
    systems.  Squares denote galaxies with SDSS spectroscopic redshift
    and $|z_{\rm cluster}-z_{\rm galaxy}|\le0.005$.  Solid gray lines
    show the Jones criteria for fossil group selection (see text).  We
    note that none of these four previously known FGs can be
    identified as fossils, if the strict Jones criteria are accepted
    (in all cases there are some galaxies inside the solid boxes,
    implying that $\Delta m_{12} < $ 2 within half of the virial
    radius).  Dashed lines show the somewhat more relaxed criteria
    used in this paper, which maintain the identification of the three
    previously known FGs as fossil systems.}
  \label{fig:magdist}
\end{figure*}

\begin{table*}
\caption{FGs Sample}\label{tab:fg}
\centering
\begin{tabular}{lccccccc}
  \hline
  \hline
  Name & $z$ & $\Delta m_{12}$& RA & Dec & Data Type & Notes & Reference\\
  \hline
  cl0245p0936 & 0.147 & 2.15 & 02:45:48.2 & +09:36:21.7 & RTT150 &new&...\\
  cl0259p0013 & 0.194 & 2.25 & 02:59:33.1 & +00:16:25.8 & SDSS &new&...\\
  cl0532m4614 & 0.135 & 1.90 & 05:32:40.3 &$-$46:11:54.4 & LCO-40 &new&...\\
  cl1038p4146 & 0.125 & 1.87 & 10:37:54.8 & +41:44:32.1 & SDSS &new&...\\
  cl1042m0008 & 0.138 & 1.80 & 10:42:17.4 &$-$00:09:34.7 & SDSS &new&...\\
  cl1110m2957 & 0.200 & 1.78 & 11:10:03.3 &$-$29:58:31.1&Danish 1.54m&new&...\\
  cl1159p5531 & 0.081 & 2.73 & 11:59:34.9 & +55:30:39.5 & SDSS&FG or
  OLEG & \cite{1999ApJ...520L...1V}\\
  cl1340p4017 & 0.171 & 2.47 & 13:40:39.3 & +40:18:25.6 & SDSS&FG or
  OLEG & \cite{1994Natur.369..462P}\\
  cl1416p2315 & 0.138 & 1.70 & 14:16:32.0 & +23:19:06.8 & SDSS& FG &
  \cite{2003MNRAS.343..627J};\\
  & & & & & & &\cite{2006AJ....132..514C};\\
  & & & & & & &\cite{2006MNRAS.369.1211K}\\
  cl2114m6800 & 0.130 & 1.99 & 21:14:24.0 &$-$68:00:55.5&Danish 1.54m&
  OLEG &\cite{1999ApJ...520L...1V}\\
  cl2220m5228 & 0.102 & 2.21 & 22:19:49.0 &$-$52:27:12.9& ENACS&new&...\\
  cl2247p0337 & 0.200 & 2.56 & 22:47:27.9 & +03:38:14.5 & RTT150 & OLEG
  & \cite{1999ApJ...520L...1V}\\
  \hline
\end{tabular}

\begin{minipage}{0.8\linewidth}
  RA and Dec are coordinates of the second brightest galaxy inside our
  search radius, $0.7r_{500}$.
\end{minipage}
\end{table*}

Figure~\ref{fig:magdist} shows our results for the four clusters
previously identified as fossil systems, using the available SDSS
data.  We show the dependence of $\Delta m_{1n}$ on the distance to
the X-ray center. Only galaxies with $m_r\le20$ are shown here. The
squares around dots indicate galaxies for which spectroscopic
redshifts are available and which belong to the group (see next
subsection). It is obvious from all four panels that the central
galaxies in these groups are isolated inside some radius, but that the
identification of the system as an FG depends critically on the search
radius. The groups cl1159p5531 and cl1340p4017 are FGs if the search
radius is equal to $0.7r_{500}$ which is slightly smaller than half
the virial radius (note that for these two groups SDSS data has not
been used in previous analyses). The group cl1416p2315 is an FG if we
relax the $\Delta m_{12}\ge2$ criterion to $\Delta
m_{12}\ge1.7$\,\footnote{The galaxy at $0.35r_{500}$ with $\Delta
  m_{1n}\approx1.7$, RA$=14^h16^m21.8^s$ and Dec$=+23^\circ17^m22.8^s$
  does not belong to the group according to its spectroscopic redshift
  (see~\citealt{2006AJ....132..514C}).}. As for the group cl1552p2013,
there are two galaxies within $0.7r_{500}$ and $\Delta m_{1n}<1.7$.
Clearly, this system had been misidentified as an FG previously (this
conclusion agrees with recent analysis by~\citealt{2008arXiv0809.2036Z}).

Based on the first three cases, we adopt a slightly relaxed optical
criterion for our sample to identify FGs in comparison to those of
J03: we classify an object as an FG if $\Delta m_{12}\ge1.7$ for
galaxies inside $0.7r_{500}$.  Almost the same criterion, $\Delta
m_{12}\ge1.75$, was used by ~\cite{2008arXiv0812.2929L}, with the aim
of optimizing the number of FGs found relative to random coincidences.
The brightest central galaxy in the whole subsample of clusters with
SDSS data has a magnitude of $m_r=13.86$.  The maximum correction for
such a galaxy is $\approx0.3$ (less correction is needed for fainter
galaxies).  We are therefore certain that we will not miss any
potential FGs due to possible errors in the photometric data.  The
search radius we choose is not much different from the assumed search
radius of $0.5r_{\rm vir}$.  Indeed, $r_{500}\approx0.6r_{\rm vir}$,
and we should take into account that $r_{500}$ found from the
$L_X-M_{500}$ correlation has its own uncertainties.  The radius $0.7r_{500}$
corresponds to the lowest $2\sigma$ boundary of $0.5r_{\rm vir}$.

\begin{figure*}
  \centerline{
    \includegraphics[width=0.33\linewidth]{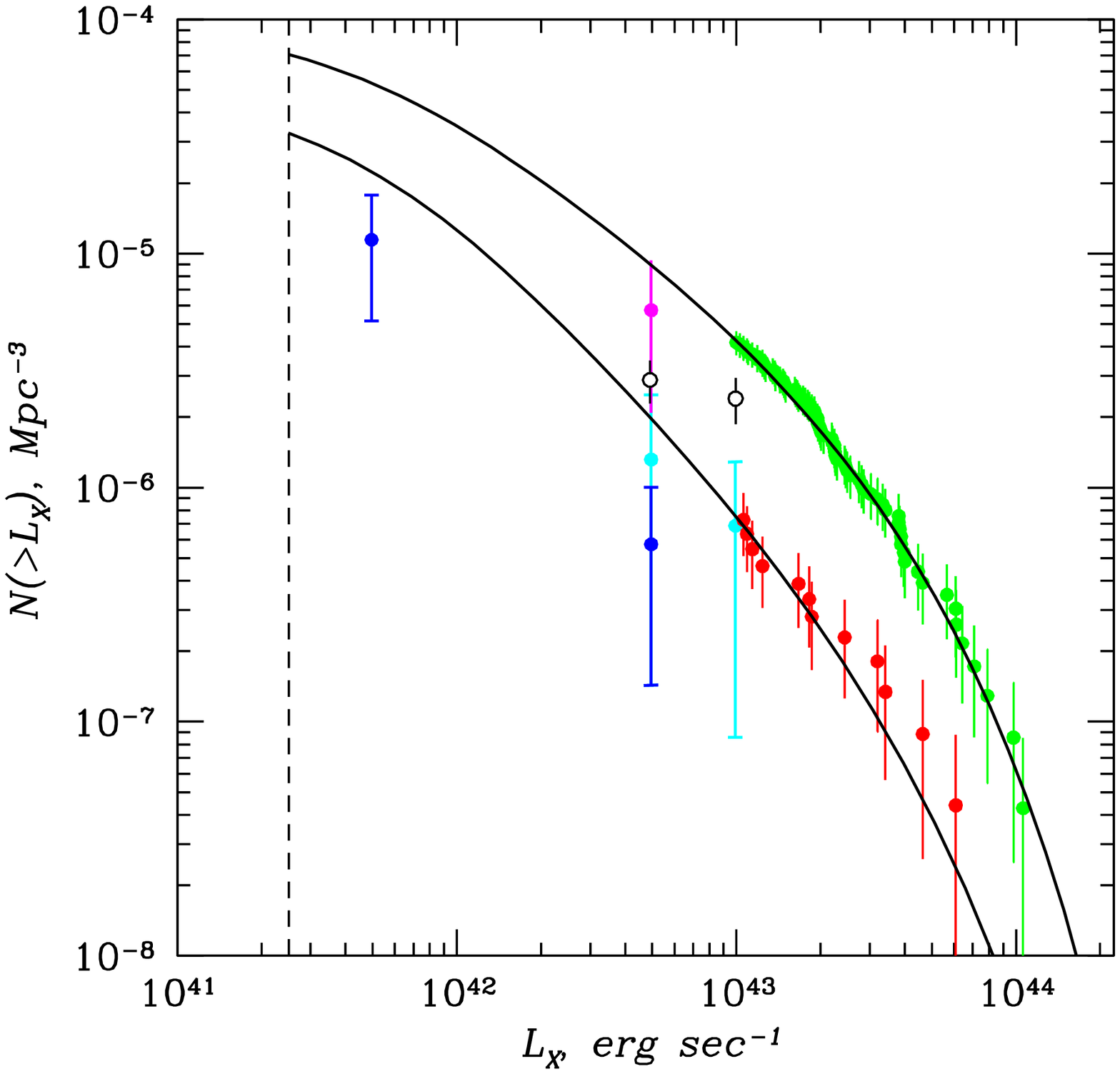}
    \includegraphics[width=0.33\linewidth]{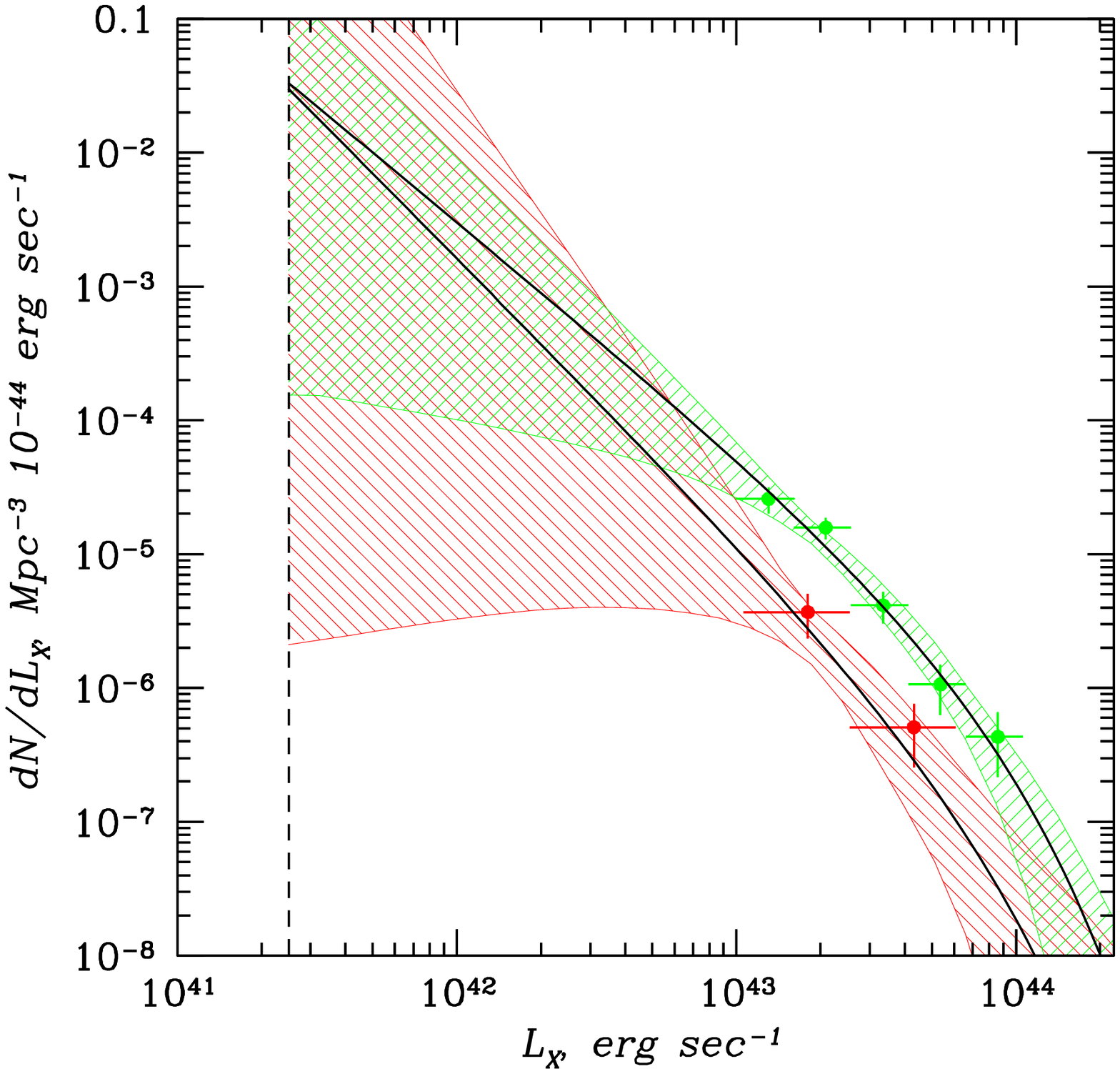}
    \includegraphics[width=0.33\linewidth]{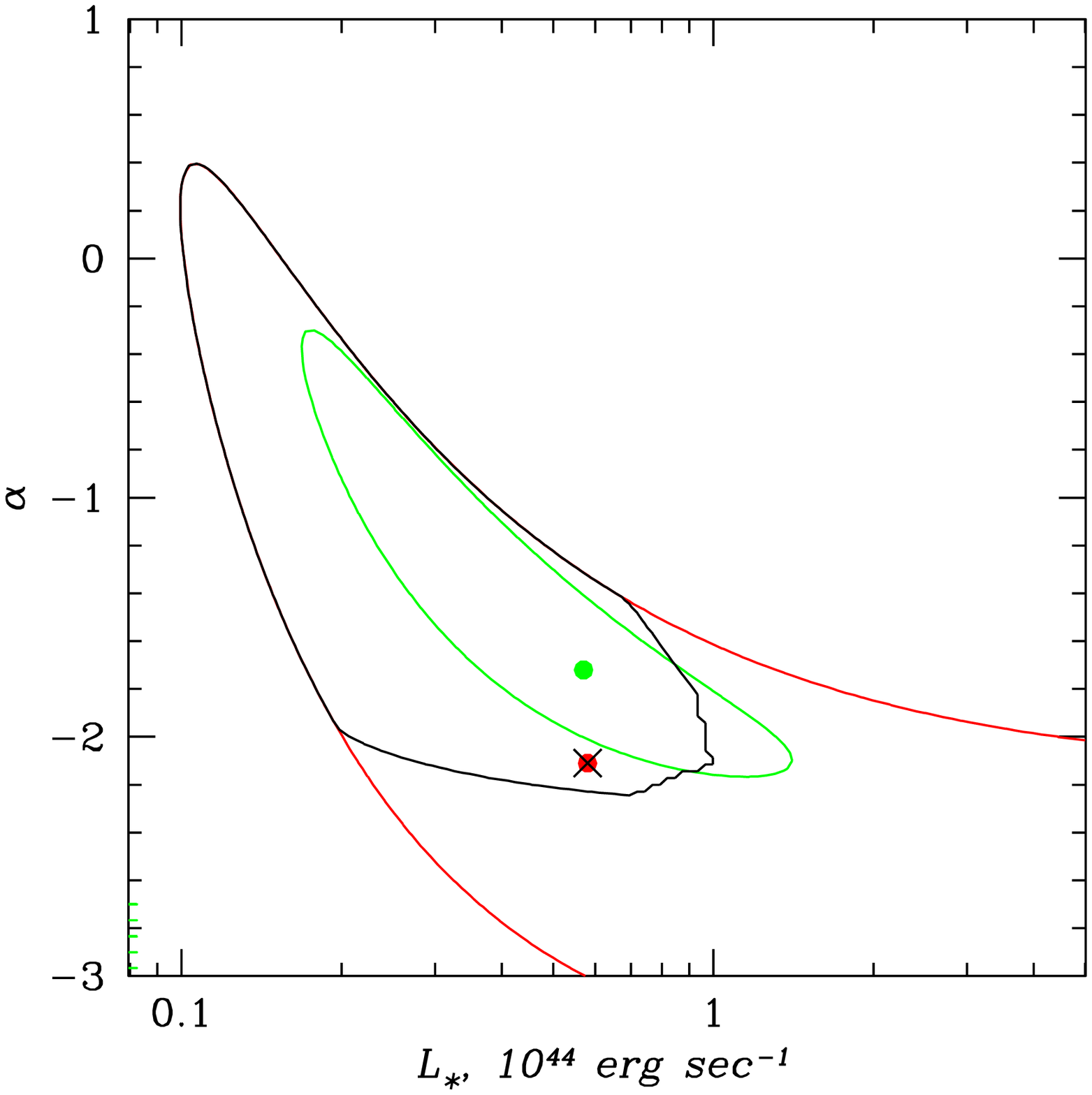} }
\caption{Luminosity functions for fossil systems.  \emph{Left panel}:
  Cumulative luminosity function. Red points correspond to fossil
  systems and green points correspond to the general population of
  clusters studied in this work.  Black lines are best fits for our
  data by the Schechter function.  Blue, cyan, magenta, and empty
  black points are previous estimations from J03,
  \cite{1998ApJ...502..558V}, \cite{2000ApJS..126..209R}, and
  \cite{2008arXiv0812.2929L} respectively.  \emph{Middle panel}:
  Differential luminosity functions for FGs (red) and clusters
  (green), studied in this work. The lines correspond to $1\sigma$
  sets of the fits obtained with different values of $L_\ast$ and
  $\alpha$ from regions defined by contours on the right panel.
  \emph{Right panel}: $1\sigma$ confidence contours on parameters of
  the Schechter function $L_\ast$ and $\alpha$ obtained with free
  normalization.  The green contour is for our whole cluster sample,
  red for FGs, and black for FGs using the constraint that the number
  density of FGs cannot be greater than the number density of
  clusters.}
\label{fig:lumfun}
\end{figure*}

\section{Search for New Fossil Groups}\label{sec3}

After having established criteria for identifying FGs from the three
previously known FGs, and setting aside the one previously
misidentified, we now analyze the remaining 71 clusters in our
sample. We divide these clusters into two sub-groups: those for which
we have optical data from the SDSS and those for which we have to rely
on optical data from other telescopes.

\subsection{Clusters with SDSS Data}

Working with SDSS data, we select galaxies with good photometry
following the recommendations from~\cite{2008ApJ...674..768O} and
applying the additional constraint that the r-filter magnitude must be
$\le21$. We also make use of SDSS spectroscopic redshifts, whenever
they are available.

We accept a given galaxy as belonging to the target cluster if
$|z_{\rm cluster}-z_{\rm galaxy}|\le0.005$, where $z_{\rm cluster}$ is
the cluster redshift cited in the 400d catalog. The selected threshold
for cluster membership, $z=0.005$, corresponds approximately to a
$2\sigma$ velocity dispersion ($1500$\,km~s$^{-1}$) for the most
massive clusters from our sample. This criterion helps us to remove
large foreground spiral galaxies from consideration.

Applying our FG criteria to clusters with SDSS data we find three new
fossil systems: cl0259p0013, cl1038p4146, and cl1042m0008.  Some of
their properties are listed in Table~\ref{tab:fg}.

\subsection{Clusters with Non-SDSS Optical Data}

In our sample there are 37 clusters for which we use R-band CCD
observations, if they are available from the 400d database. We use the
SExtractor program to perform the analysis of the CCD plates. We set
the detection threshold equal to $4\div5\sigma$ over the background
level and in order to be consistent with SDSS photometry, we measure
the galaxy fluxes in Petrosian apertures. In the left panel of
Figure~\ref{fig:photom} the good agreement between the SDSS and our
photometries is demonstrated. Larger discrepancies only arise for
faint galaxies, which are not important for our study. We note that
the SDSS archival measurements are done in the r-filter, while our own
measurements are carried out in the R-filter. This leads to different
apparent magnitudes.  However, due to the linear transformations
between the r and the R~filter~\citep{1996AJ....111.1748F} this
difference does not affect our $\Delta m_{1n}$ measurements strongly
(see also the left panel of Figure~\ref{fig:photom}). Applying our FG
criteria to the data we find four more new FGs: cl0245p0936,
cl0532m4614, cl1110m2957, and cl2220m5228. For the cluster cl2220m5228
we do not have a CCD image from the 400d database, but we found that
it is a member of the ENACS survey~\citep{1998A&AS..129..399K}.  Using
ENACS data we can establish that this cluster is an FG (see Appendix
for details.)

To summarize, we detect seven new FGs in the 400d catalog. Besides
cl1552p2013, all previously known FGs satisfy our criteria.
Table~\ref{tab:fg} shows the results from all twelve FGs from our
sample. In the table we note a previous identification of the object
as a fossil group or OLEG, its redshift, and $\Delta m_{12}$.  We also
give coordinates of the second brightest galaxy. In the Appendix we
provide a more detailed description of each system.

\section{Number Density of Fossil Groups}\label{sec4}

\begin{table}
  \caption{Parameters for the best fit Schechter functions.}
  \label{tab:fit_par}
  \centering
  \medskip\def\arraystretch{1.15}
  \begin{tabular}{clcc}
    \hline
    \hline
    Sample & $A$, Mpc$^{-3}$ $10^{-44}$ erg~sec$^{-1}$& $L_\ast$,
    $10^{44}$ erg sec$^{-1}$& $\alpha$\\
    \hline
    FG  & $(3.20\pm1.61)\times10^{-7}$ & 0.58& $-2.11$\\
    ALL & $(2.87\pm0.56)\times10^{-6}$ & 0.57& $-1.72$\\
    \hline
  \end{tabular}

  \begin{minipage}{0.8\linewidth}
    Normalization errors are obtained by marginalizing over
    $L_\ast$ and $\alpha$.
  \end{minipage}
\end{table}

There are several estimates for the number density of FGs in the
literature~\citep{1998ApJ...502..558V,
  2000ApJS..126..209R,2003MNRAS.343..627J,2007AJ....134.1551S,2008arXiv0812.2929L}.
Our sample includes both known and newly identified FGs in the 400d
catalog. While our definition for FGs has been derived using only
systems for which SDSS data is available, nevertheless it also works
well for other previously known FGs without SDSS data: no fossil
systems previously detected in the 400d survey were missed.
Consequently, we can use our sample to derive a constraint on the
number density of FGs as a function of their X-ray luminosity.

In order to find the number density of objects with a given luminosity
(luminosity function) we need to know the corresponding survey volume.
The 400d survey is a flux limited survey, and therefore we can use
standard techniques for the calculation of the volume, where all
clusters of a given luminosity would have been detected. The volume is
given by (see~\citealt{2007ApJS..172..561B} for details):
\begin{equation}
V(L) = \int_{0.0032}^{0.20}A(f_X,z)\frac{dV}{dz}dz,
\end{equation}
where the lower integration limit corresponds to the 400d cluster with
the lowest redshift, the upper integration limit is our limit for
sample selection, and the function $A(f_X,z)$ is an effective survey
area (see the 400d survey data). We fit our FG sample by a Schechter
function~\citep{1976ApJ...203..297S} (the use of the Schechter
function is justified below):
\begin{equation}\label{eq2}
  f(L)dL = A \left(\frac{L}{L_\ast}\right)^\alpha
  \exp\left(-\frac{L}{L_\ast}\right)dL,
\end{equation}
giving the number of clusters with luminosities from $L$ to $L+dL$
inside a unit volume. We use the maximum likelihood method for
unbinned data~\citep{1979ApJ...228..939C}, with the best-fit
parameters found by maximizing the likelihood function:
\begin{equation}
\ln\pounds  = \sum_i \ln(f(L_i)V(L_i)) - \int f(L)V(L)dL.
\end{equation}
The cumulative luminosity function for our FGs is shown in the left
panel of Figure~\ref{fig:lumfun} by the red points and the best fit is
shown by the black line (the parameters for the fit are given in
Table~\ref{tab:fit_par}).

We also fit the luminosity functions for the entire cluster sample
under consideration (75 clusters) via a Schechter function. The
results are shown by the green points in the left panel in
Figure~\ref{fig:lumfun}. The best fit is shown by the black line (for
the best-fit parameters, see Table~\ref{tab:fit_par}). In the middle
panel of Figure~\ref{fig:lumfun} we show the differential luminosity
functions for FGs and clusters.

For comparison with previous results we extrapolate the FG and cluster
number density fits to $L_X=2.5\times10^{41}$ erg~sec$^{-1}$ which
corresponds to the group
boundary~\citep{2001MNRAS.328..461O,2003MNRAS.343..627J} -- beyond
this value objects are more likely to be isolated elliptical galaxies than
groups. We use Table~2 from J03 as a compilation of known results. The
points from this table, rescaled to the cosmology used here, are shown
in the left panel of Figure~\ref{fig:lumfun}.  Blue points correspond to
the estimates obtained by J03, cyan correspond to estimates
from~\cite{1999ApJ...520L...1V}, the magenta point is an estimate
from~\cite{2000ApJS..126..209R}, and the empty black points are
estimates from the recent work of~\cite{2008arXiv0812.2929L}.

Overall, our constraints are in good agreement with previous results.
The FG number density of~\cite{2000ApJS..126..209R} (magenta point) is
most likely overestimated, but their results are based on only 3
objects and therefore may be considered to be consistent with the
other findings, given the uncertainties.  Our results are in good
agreement with the number density estimation of J03, and in perfect
agreement with the estimate from~\cite{1999ApJ...520L...1V}. The
latter agreement is not surprising since their data set is a subset of
ours, and the same methodology was used for survey area
calibration. The number density estimation
from~\cite{2008arXiv0812.2929L} agrees with the extrapolation of our
result, but it is too high for FGs brighter than $10^{43}$
erg~s$^{-1}$. We note that the search radius, 350~kpc, chosen in that
work is much smaller than the half or virial radius for such X-ray
bright objects, possibly causing some misidentifications and an
overall overestimation of the FG number density.

Despite the increased number of FGs in comparison to previous
results, our constraints are still not very tight.  In the right panel
of Figure~\ref{fig:lumfun} we show the $1\sigma$ confidence contours
for $\alpha$ and $L_\ast$ obtained with a free normalization parameter
(red represents FGs and green, clusters). In the middle panel of
Figure~\ref{fig:lumfun} we show the sets of fits corresponding to the
contours on the right panel via red and green shaded regions. We can
reduce the range of allowed values for $L_\ast$ and $\alpha$ for the
FG fit by taking into account that the FG number density cannot be
higher than the number density of clusters (at least in the luminosity
range where the phrase ``group of galaxies'' is still valid, i.e.,
where $L_X\ge2.5\times10^{41}$ erg~sec$^{-1}$).  We use this
constraint in our fitting procedure.  The best fit remains the same
with and without it, but the $1\sigma$ intervals for $\alpha$ and
$L_\ast$ become tighter (black contour in the right panel of
Figure~\ref{fig:lumfun}).

\subsection{Schechter Function for Fossil Groups}

As can be seen from the left and middle panels of
Figure~\ref{fig:lumfun}, the Schechter function provides a very good fit
to the cluster luminosity function. This is not surprising since even
a simple theory of large scale structure formation
~\citep{1974ApJ...187..425P} predicts Schechter-like functions.  Since
the cluster luminosity is connected with the cluster mass via a
power-law relation and the range of luminosities considered here
covers only one decade of luminosities, Eq.~(\ref{eq2}) has
enough degrees of freedom to fit the mass function or the luminosity
function in our case. Of course, our extrapolation beyond the data
range should be treated with caution, but we do this mainly to compare
our results with previous studies.

We fit the FG luminosity function with a Schechter function as well.
The simple justification is that FGs are a subset of the cluster
sample, and the shape of their luminosity function should be similar
to the shape of the cluster luminosity function. A more precise result
for the FG number density may be obtained by using the extended
Press-Schechter formalism~\citep{2006ApJ...637L...9M}, but for a
sample of only 12 objects this is hardly necessary.

\begin{figure}
  \plotone{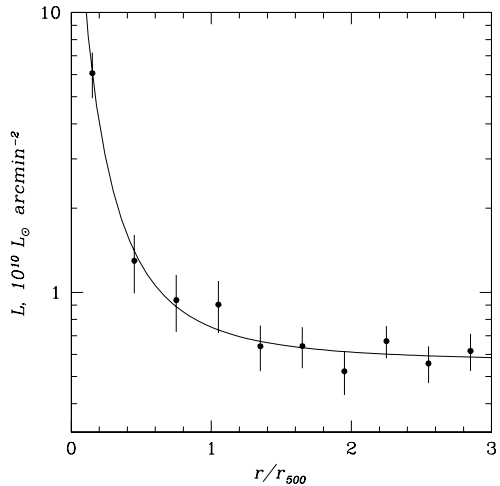}
  \caption{Light profile of cluster cl1629p2123 and a projected NFW
    profile plus constant background fit (solid line).}
 \label{fig:nfw}
\end{figure}

\section{$L_X-L_r$ Relation}\label{sec5}

It was shown in J03 and~\cite{2007MNRAS.377..595K} that for a given
X-ray luminosity, FGs are less luminous in the optical band as
compared to (X-ray bright) normal groups of galaxies.  However, these
comparisons were not uniform, i.e., the data for FGs and other systems
were obtained separately, using different instruments and methods of
data analysis.  Our data comes from a uniform sample, and we apply the
same data analysis approach to the total sample of systems. We are
therefore well positioned to study X-ray and optical luminosities for
different systems.  For this study we use clusters with SDSS data to
measure optical luminosities.  As elsewhere in this paper, we use
X-ray luminosities from the 400d catalog.

\subsection{Measurements of Optical Luminosities}

We use SDSS data to select galaxies around the X-ray centers of the
clusters. We select all galaxies brighter than $m_r^{\rm th}=21$ and
require the photometry flags to be the same as in Section~\ref{sec2}.
After that we carry out the following steps~\citep{mescheryakov}:

\begin{table*}
  \def\tnote#1{\ensuremath{^{\text{#1}}}}
  \def\d{\phantom{1}}
    \centering
    \caption{Optical luminosities of clusters with SDSS data}\label{tab:sample}
    \medskip\def\arraystretch{1.15}
    \footnotesize
    \begin{tabular}{lccrccc}
      \hline
      \hline
      \multicolumn{1}{c}{Name} & $z$ & $L_X$ & $r_{500}$ &
      $L_r(r<r_{500})$ & $M_{500}$\\ 
       & & $10^{43}$ erg s$^{-1}$ & kpc & $10^{11}L_\odot$ & $10^{14}M_\odot$\\
      \hline
cl0050m0929  & 0.199 &  3.81 & 784 & $5.69_{-0.76}^{+2.07}$ & 1.72\\ 
cl0223m0852  & 0.163 &  1.73 & 679 & $3.39_{-0.63}^{+1.70}$ & 1.07\\ 
cl0259p0013\tnote{fg}  & 0.194 &  3.19 & 757 & $3.71_{-0.44}^{+2.14}$ & 1.54\\ 
cl0304m0702  & 0.135 &  1.19 & 638 & $1.16_{-0.29}^{+0.72}$ & 0.86\\ 
cl0810p4216  & 0.064 &  2.24 & 754 & $3.03_{-0.17}^{+1.34}$ & 1.33\\ 
cl0838p1948  & 0.123 &  1.96 & 712 & $1.58_{-0.50}^{+1.93}$ & 1.19\\ 
cl0907p1639  & 0.076 &  1.98 & 731 & $4.10_{-0.37}^{+1.67}$ & 1.23\\ 
cl0910p6012\tnote{*}  & 0.181 &  1.24 & 627 & $1.07_{-0.20}^{+0.20}$ & 0.86\\ 
cl0943p1644  & 0.180 &  1.80 & 678 & $3.71_{-0.48}^{+1.48}$ & 1.09\\ 
cl1013p4933  & 0.133 &  1.95 & 707 & $3.36_{-0.52}^{+1.53}$ & 1.18\\ 
cl1038p4146\tnote{fg}  & 0.125 &  1.06 & 626 & $3.09_{-0.29}^{+1.17}$ & 0.81\\ 
cl1042m0008\tnote{fg}  & 0.138 &  1.67 & 683 & $2.06_{-0.24}^{+1.09}$ & 1.07\\ 
cl1124p4155  & 0.195 &  3.97 & 792 & $2.13_{-0.60}^{+1.58}$ & 1.76\\ 
cl1142p1008  & 0.119 &  1.63 & 687 & $2.41_{-0.42}^{+0.98}$ & 1.06\\ 
cl1142p1027  & 0.117 &  1.10 & 633 & $1.85_{-0.22}^{+0.75}$ & 0.83\\ 
cl1142p2145  & 0.131 &  1.95 & 708 & $4.95_{-0.46}^{+0.46}$ & 1.18\\ 
cl1146p2854  & 0.149 &  2.19 & 718 & $3.07_{-0.43}^{+1.07}$ & 1.25\\ 
cl1159p5531\tnote{fg}  & 0.081 &  1.14 & 650 & $2.97_{-0.23}^{+2.38}$ & 0.87\\ 
cl1212p2727  & 0.179 &  6.42 & 883 & $5.55_{-0.47}^{+1.64}$ & 2.40\\ 
cl1217p2255  & 0.140 &  1.48 & 665 & $1.86_{-0.22}^{+0.73}$ & 0.99\\ 
cl1222p2559  & 0.160 &  1.26 & 637 & $3.17_{-0.52}^{+1.35}$ & 0.88\\ 
cl1231p4137  & 0.176 &  2.26 & 712 & $2.43_{-0.46}^{+1.23}$ & 1.26\\ 
cl1235p4117  & 0.189 &  2.81 & 740 & $3.27_{-0.63}^{+0.63}$ & 1.43\\ 
cl1236p2550  & 0.175 &  2.01 & 696 & $2.77_{-0.59}^{+0.59}$ & 1.17\\ 
cl1340p3958  & 0.169 &  2.55 & 733 & $4.35_{-0.45}^{+0.45}$ & 1.36\\ 
cl1340p4017\tnote{fg}  & 0.171 &  1.24 & 631 & $1.38_{-0.78}^{+1.84}$ & 0.87\\ 
cl1341p2622  & 0.075 & 10.60 &1030 & $7.18_{-0.47}^{+2.76}$ & 3.48\\ 
cl1349p4918  & 0.167 &  3.88 & 801 & $4.77_{-0.77}^{+1.87}$ & 1.77\\ 
cl1416p2315\tnote{fg}  & 0.138 &  6.09 & 893 & $4.51_{-0.49}^{+2.59}$ & 2.38\\ 
cl1436p5507  & 0.125 &  1.03 & 622 & $1.36_{-0.36}^{+1.09}$ & 0.80\\ 
cl1438p6423  & 0.146 &  1.43 & 659 & $3.58_{-0.45}^{+0.45}$ & 0.96\\ 
cl1515p4346  & 0.137 &  1.62 & 679 & $2.14_{-0.23}^{+0.65}$ & 1.05\\ 
cl1533p3108  & 0.067 &  1.90 & 728 & $3.68_{-0.33}^{+0.64}$ & 1.20\\ 
cl1537p1200  & 0.134 &  1.19 & 638 & $3.89_{-0.47}^{+1.19}$ & 0.87\\ 
cl1552p2013  & 0.136 &  2.29 & 730 & $5.13_{-0.46}^{+2.19}$ & 1.30\\ 
cl1629p2123  & 0.184 &  2.24 & 708 & $4.14_{-0.76}^{+0.76}$ & 1.25\\ 
cl1630p2434  & 0.066 &  1.76 & 717 & $2.39_{-0.19}^{+0.63}$ & 1.15\\ 
cl1639p5347  & 0.111 &  3.84 & 823 & $3.37_{-0.51}^{+1.85}$ & 1.82\\
      \hline
    \end{tabular}
\begin{minipage}{0.8\linewidth}
  $^{\rm fg}$ -- fossil group\\ $^*$ -- Due to a poor NFW fit, the
  luminosity is estimated in these cases as a sum of galaxy
  luminosities inside a given radius minus the sum of luminosities of
  background galaxies corrected for the area. Lost light corrections
  are also applied.\\
\end{minipage}
\end{table*}

\begin{itemize}

\item[1)] Exclude central cD galaxies.

\item[2)] Mask regions around bright stars and big foreground
  galaxies, because the photometry around such objects may be incorrect,
  and the photometry flags usually do not select these galaxies. We also
  do not consider galaxies with SDSS spectroscopic redshifts if
  $|z_{\rm cluster} - z_{\rm galaxy}|>0.005$.

\item[3)] Estimate the lost light due to the selected magnitude
  threshold.  We subtract the background luminosity function from the
  cluster luminosity function, where the cluster luminosity function
  is built for galaxies around the X-ray center within a radius of
  $0.5r_{500}$, and a background function is built for galaxies inside
  the ring with radii from $2r_{500}$ to $3r_{500}$. We fit the
  residual by a Schechter function written in the form:
  \begin{equation}
    \begin{split}
      \phi(m)dm = \phi_0 10^{0.4(m_\ast-m)(\alpha+1)} \\
      \exp(-10^{0.4(m_\ast-m)})dm.
    \end{split}
  \end{equation}
   The fraction of the lost light is given by:
  \begin{equation}
    f = \Gamma(\alpha+2,10^{0.4(m_\ast-m^{\rm th})})/\Gamma(\alpha+2).
  \end{equation}

\item[4)] Transform magnitudes of all galaxies inside $3r_{500}$ to
  luminosities with $K$-corrections (kcorrect v~4.1.4
  \citep{2007AJ....133..734B}) for cluster redshift. Build the light
  profile by summing luminosities of individual galaxies inside the
  rings around the cluster center and exclude regions masked in step
  2).

\item[5)] Fit the light profile by a projected Navarro-Frenk-White
  function~\citep{1997ApJ...490..493N} plus constant background.  An
  example of such a fit is shown in Figure~\ref{fig:nfw}.

\item[6)] In order to calculate the total cluster luminosity inside a
  given radius we integrate the NFW fit over the volume, then make
  corrections for the lost light (see step 3) and add the luminosity
  of cD galaxies.  The error for the cD luminosities takes into
  account an underestimation of the measured magnitude due to the SDSS
  background subtraction algorithm. Therefore, we increase the upper
  luminosity error by the appropriate amount (see Figure~3
  in~\citealt{2008ApJS..175..297A}). Our measurements are given in
  Table~\ref{tab:sample}.

\end{itemize}

\begin{figure}
  \plotone{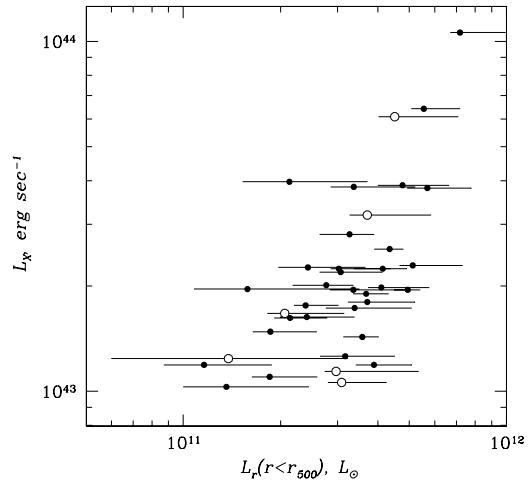}
  \caption{X-ray luminosity vs. optical luminosity measured inside
    $r_{500}$. Empty points show FGs.}
 \label{fig:Lx_Lr}
\end{figure}

\begin{figure}
  \plotone{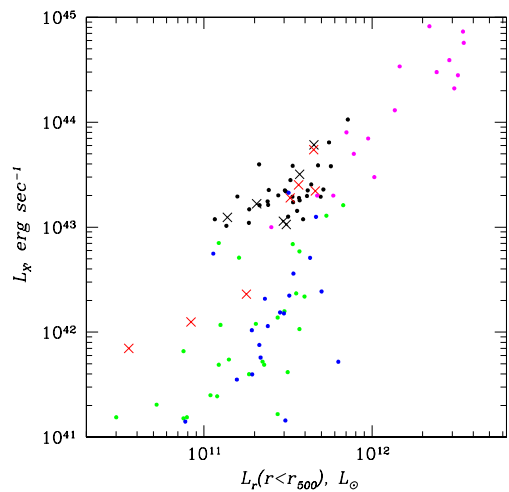}
  \caption{Comparison of our measurements with previous studies. 
  Black dots and black crosses correspond to clusters and FGs from this
  work, red crosses show FGs from~\cite{2007MNRAS.377..595K}, blue
   and green dots represent groups studied in~\cite{2003MNRAS.340..485H}
    and~\cite{2004MNRAS.350.1511O} correspondingly. Magenta dots
    correspond to clusters of galaxies from~\cite{2002AstL...28..366V}.}
 \label{fig:comp}
\end{figure}

We show the $L_X-L_r(r<r_{500})$ correlation in Fig.~\ref{fig:Lx_Lr},
where empty points correspond to FGs.  We find that for a given
optical luminosity, the FGs in our sample are not systematically
brighter in the X-ray than other clusters.  Indeed, four of the FGs
reside on the upper envelope of the correlation, but there are also
two FGs (cl1038p4146 and cl1159p5531) which appear X-ray faint
compared to other clusters with similar optical luminosity.

We also compare our results with previous studies.
~\cite{2007MNRAS.377..595K} compared their FGs with other groups of
galaxies studied by \cite{2003MNRAS.340..485H} and
\cite{2004MNRAS.350.1511O}.  We show these data in
Fig.~\ref{fig:Lx_Lr}.  We also add a compilation of optical luminosity
measurements from~\cite{2002AstL...28..366V}, with optical
luminosities rescaled to $r_{500}$. There appears to be general
agreement in the trends between the FGs studied here and the FGs
studied in \cite{2007MNRAS.377..595K}; their distribution is also in
agreement with the overall data for clusters, but not for groups.  It
is not clear whether the difference between FGs and other groups
reported by ~\cite{2007MNRAS.377..595K} is real for less massive
systems or if it is a result of systematic differences in analysis
techniques used in different studies.  We hope to address this issue
in future work.

\section{Summary}\label{sec6}

We have analyzed nearby ($z\le0.2$) and X-ray bright ($L_X \ge
10^{43}$ erg~sec$^{-1}$) clusters of galaxies from the 400d catalog.
By evaluating known FGs against SDSS data, we have formulated slightly
revised selection criteria for fossil systems. Our criteria are
somewhat relaxed compared to those used by J03, but their advantages
are: 1) they select all previously known FGs (except for one system
clearly misidentified as an FG), 2) they account for the possibility
that the photometry of the brightest galaxy may be underestimated and
therefore guard against missing any FGs as a consequence.

Our main results can be summarized as follows:

1) We found seven new FGs in the 400d cluster survey. The images and
descriptions of all 400d FGs are given in Appendix A.

2) We put new constraints on the number density of FGs. Our
constraints are consistent with ones obtained previously, but are
tighter due to the larger number of FGs studied.

3) We measured optical luminosities of clusters with SDSS photometric
data. Measurements of $L_X-L_r$ correlations for these clusters show
that FGs are similar to the overall cluster sample, and follow the
scaling relation of clusters of galaxies.

Current definitions of FGs are not robust, i.e., the same object may
be identified as an FG or not depending on what magnitude threshold
and search radius is chosen.  Due to the variety of galaxy
distributions in cluster potential wells, and due to cluster
evolution, it is hard to motivate a strict definition of FGs using
simple observables.  Moreover, there is another major obstacle from
the observational perspective for FG searches -- the group
membership, which cannot be firmly established observationally. While
precise redshift measurements can weed out most of the projected
galaxies, they are not enough to establish the presence or absence of
a gravitational bond between a given galaxy and the group, nor can
they provide a measurement of the distance from a given galaxy to the
center along the line of sight. It is evident that the situation is
much more complicated if precise redshift measurements are not
available.  The absence of such redshift measurements may account for
the differences in the number density of FGs found in observations and
in simulations (see e.g.~\citealt{2007MNRAS.382..433D}).  Most
observational estimates for the number density of FGs most likely
represent a lower limit for the actual number.

Due to the small number of known FGs it is too early to draw final
conclusions about their properties. Any new FG candidate must be
examined thoroughly.  Optical selection of FGs is possible, but
complicated~\citep{2008ApJ...684..204V}. Dedicated X-ray observations
would be very useful to continue studies of unusual properties of FGs.

\acknowledgements

This work was supported by the LDRD program at Los Alamos National
Laboratory. Data from SDSS DR6 was used extensively in this study. The
SDSS is managed by the Astrophysical Research Consortium for the
Participating Institutions. The Participating Institutions are the
American Museum of Natural History, Astrophysical Institute Potsdam,
University of Basel, University of Cambridge, Case Western Reserve
University, University of Chicago, Drexel University, Fermilab, the
Institute for Advanced Study, the Japan Participation Group, Johns
Hopkins University, the Joint Institute for Nuclear Astrophysics, the
Kavli Institute for Particle Astrophysics and Cosmology, the Korean
Scientist Group, the Chinese Academy of Sciences (LAMOST), Los Alamos
National Laboratory, the Max-Planck-Institute for Astronomy (MPIA),
the Max-Planck-Institute for Astrophysics (MPA), New Mexico State
University, Ohio State University, University of Pittsburgh,
University of Portsmouth, Princeton University, the United States
Naval Observatory, and the University of Washington. We acknowledge
useful discussions regarding SDSS photometry with Chris Miller and
Adrian Pope.  This research has made use of the NASA/IPAC
Extragalactic Database (NED) which is operated by the Jet Propulsion
Laboratory, California Institute of Technology, under contract with
the National Aeronautics and Space Administration.

\pagebreak
\section*{Appendix}
\label{app}

\subsection*{Description of the Fossil Groups}

In this appendix we provide a detailed description of all twelve FGs
studied in this paper. This information will be useful for future FG
studies. For example, higher quality data for the systems studied here
might become available or one might want to refine the definition of
FGs further. In either case any detailed information available
regarding these FGs will be helpful.

In all images the inner black circle marks the radius $0.7r_{500}$,
which we have adopted as our search radius. The outer black circle
shows the radius $r_{500}$, and a white cross marks the X-ray center.
The white arrow inside $0.7r_{500}$ points to the second brightest
galaxy in the group.  The second white arrow inside the ring
$0.7r_{500}$ -- $r_{500}$ (if present) shows the second brightest
galaxy within the increased search radius. Black arrows show projected
galaxies with known spectroscopic redshifts, which do not belong to
the FG.  Such galaxies could have led to an incorrect classification
of the groups as being non-fossil, if the redshift measurements were
not available.

\begin{figure}
  \plotone{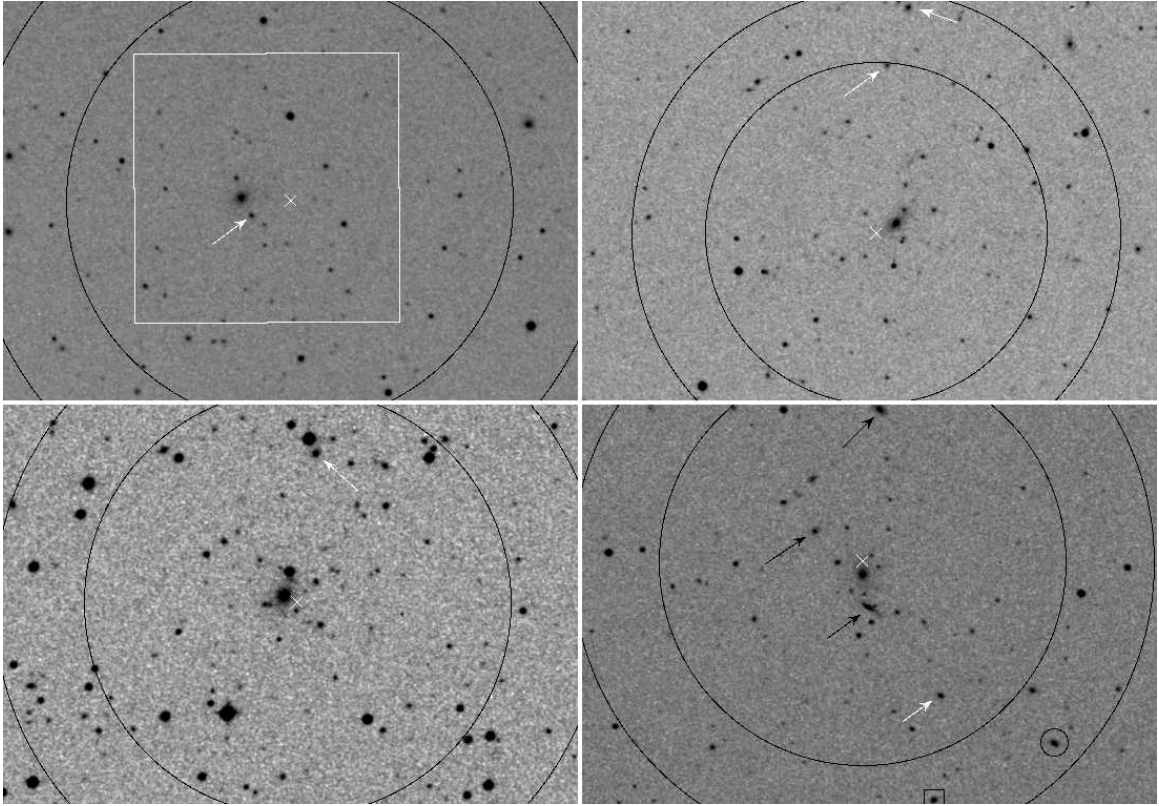}
  \caption{DSS2 images of FGs. From left to right, top to bottom:
    cl0245p0936, cl0259p0013, cl0532m4614, and cl1038p4146.}
 \label{fig:fg1234}
\end{figure}

\subsubsection*{cl0245p0936}

cl0245p0936 is one of the new fossil systems found by us in the 400d
survey at $z=0.147$. The RTT150 CCD unfortunately does not cover a
full circle with radius $0.7r_{500}$. The coverage is shown as a white
square in Figure~\ref{fig:fg1234}. The second brightest galaxy is
marked by the white arrow, $\Delta m_{12}=2.15$. A visual inspection
of the DSS2 image reveals that inside $0.7r_{500}$ there are no
galaxies brighter than the second brightest.  However, within
$r_{500}$ there may be galaxies which are brighter than the second
brightest. It is not possible to conclude from the currently available
data whether those galaxies belong to the system or not.

\subsubsection*{cl0259p0013}

cl0259p0013 is another new fossil system discovered by us at
$z=0.194$. For this object, optical data from the SDSS is available.
The second brightest galaxy is marked with a white arrow in
Figure~\ref{fig:fg1234} with $\Delta m_{12}=2.25$. If we increase the
search radius up to $r_{500}$ the second brightest galaxy will be at
RA=$02^h59^m31.704^s$ Dec=$+00^\circ17^m21.38^s$ (another white arrow
) for which $m_{12}=1.92$, i.e. this group would satisfy our magnitude
threshold, even with the increased search radius.

\subsubsection*{cl0532m4614}

cl0532m4614 is the third newly identified fossil system. The object
is at $z=0.135$. The optical image was obtained on the LCO-40
telescope. The second brightest galaxy is shown by a white arrow in
Figure~\ref{fig:fg1234} with $\Delta m_{12}=1.90$.  If we increase the
search radius up to $r_{500}$ we do not find any galaxy brighter than
the second brightest galaxy and therefore the object would be a fossil
group even with a larger search radius.

\subsubsection*{cl1038p4146}

cl1038p4146 is the fourth of our new fossil systems. For this object,
which is at $z=0.125$, SDSS data is available. The second brightest
galaxy, for which the redshift is known, is marked by a white arrow in
the Figure~\ref{fig:fg1234} with $\Delta m_{12}=1.87$. The image shows
three bright projected galaxies (black arrows in
Figure~\ref{fig:fg1234}) for which redshifts are fortunately
available. Without these redshifts, the object would not have been
classified as an FG.  Increasing the search radius up to $r_{500}$
reveals two bright galaxies belonging to the cluster, which would not
allow us to classify the system as an FG. These galaxies have
coordinates $10^h37^m44.679^s$ $+41^\circ43^m49.89^s$, $\Delta
m_{12}=1.12$ (small black circle on the Figure~\ref{fig:fg1234}) and
$10^h37^m54.881^s$ $+41^\circ42^m49.86^s$ $\Delta m_{13}=1.41$ (small
black box on the Figure~\ref{fig:fg1234}).

\begin{figure}
  \plotone{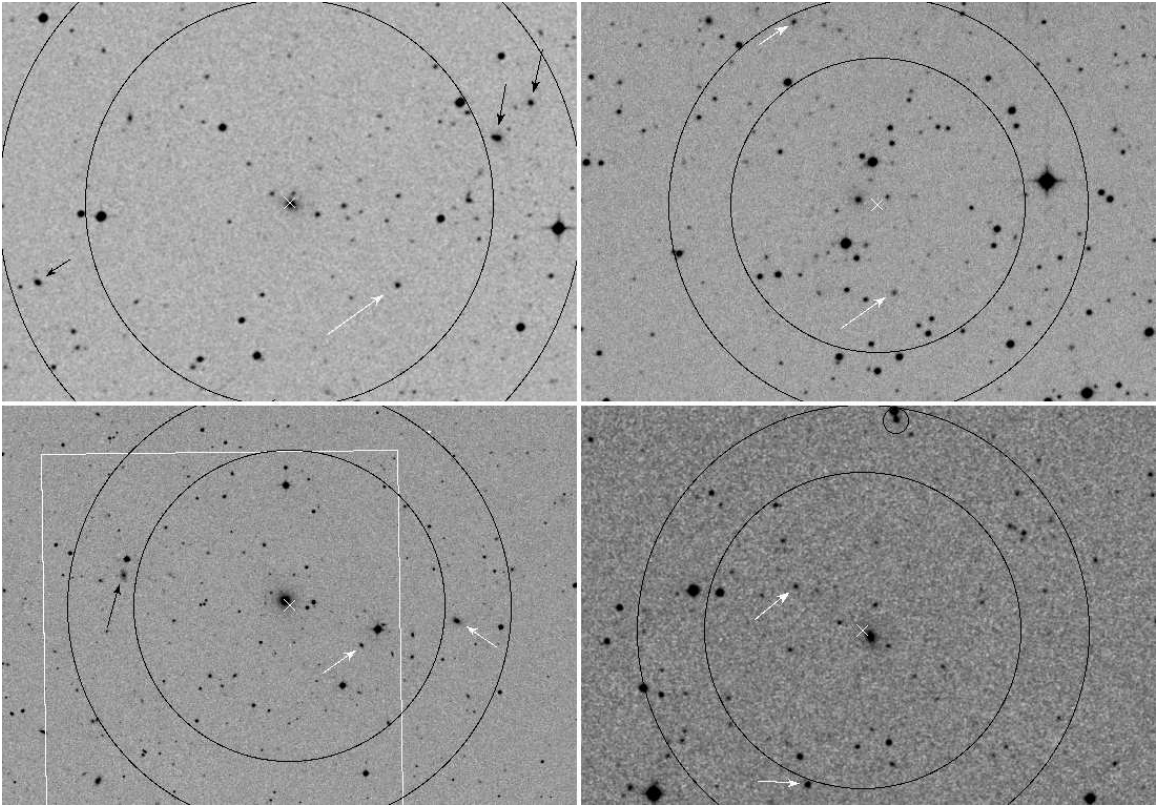}
  \caption{DSS2 images of FGs. From left to right top to
    bottom: cl1042m0008, cl1110m2957, cl1159p5531, and cl1340p4017.}
 \label{fig:fg5678}
\end{figure}

\subsubsection*{cl1042m0008}

cl1042m0008 is another newly discovered fossil system. SDSS data is
available for this object, which is at $z=0.138$. The second brightest
galaxy is shown by a white arrow in Figure~\ref{fig:fg5678} with
$\Delta m_{12}=1.80$. Increasing the search radius up to $r_{500}$
does not change its FG identification (the second brightest galaxy
remains the same). More bright galaxies appear, but they are all
projected according to their spectroscopic redshifts. Their positions
are indicated by black arrows in Figure~\ref{fig:fg5678}.

\subsubsection*{cl1110m2957}

cl1110m2957 is the sixth new FG. For this object, which is at
$z=0.20$, data from the Danish 1.54-m telescope is available. The
second brightest galaxy is marked by a white arrow (see
Figure~\ref{fig:fg5678}) with $\Delta m_{12}=1.78$.  With an increase
of the search radius to $r_{500}$, the object would no longer be
classified as an FG.  The new bright galaxy is located at
RA=$11^h10^m10.667^s$ Dec=$-29^\circ54^m08.92^s$, with $m_{12}=1.32$,
and is shown by another white arrow on the
Figure~\ref{fig:fg5678}. The redshifts for both second brightest
galaxies are unknown.

\subsubsection*{cl1159p5531}

cl1159p5531 has been previously identified as an
OLEG~\cite{1999ApJ...520L...1V} using X-ray data from \emph{ROSAT}
PSPC pointed observation and optical data from the FLWO2 telescope
(the size of CCD image is shown by the white rectangle in
Figure~\ref{fig:fg5678}).  This object has SDSS data and we also used
it as a reference to re-define the FG criteria. The second brightest
galaxy inside $0.7r_{500}$ with $\Delta m_{12}=2.73$ has a measured
spectroscopic redshift and belongs to the cluster (see white arrow in
Figure~\ref{fig:fg5678}), which is at $z=0.081$. If we increase the
search radius, this group may still be considered an FG (another
bright galaxy has $\Delta m_{12}=1.69$, it belongs to the group, see
the white arrow in Figure~\ref{fig:fg5678}, and see also the first
panel in Figure~\ref{fig:magdist}), even though $\Delta m_{12}$ would
be slightly smaller than our threshold.

\subsubsection*{cl1340p4017}

cl1340p4017 is the first ever identified fossil
group~\citep{1994Natur.369..462P}.  SDSS data is now available for
this object, which is at $z=0.171$, and we used it to re-define the FG
criteria (see second panel in Figure~\ref{fig:magdist}). The second
brightest galaxy inside a search radius $0.7r_{500}$ is shown by a
white arrow and has $\Delta m_{12}=2.47$.  Increasing the search
radius to $r_{500}$ reveals two more bright galaxies (see the white
arrow inside the ring in Figure~\ref{fig:fg5678} for a galaxy with
known spectroscopic redshift and $\Delta m_{12}=1.31$, and a black
circle for a galaxy with only photometric data and $\Delta
m_{12}=1.58$). With these galaxies included, this object would not be
classified as an FG.

\begin{figure}
  \plotone{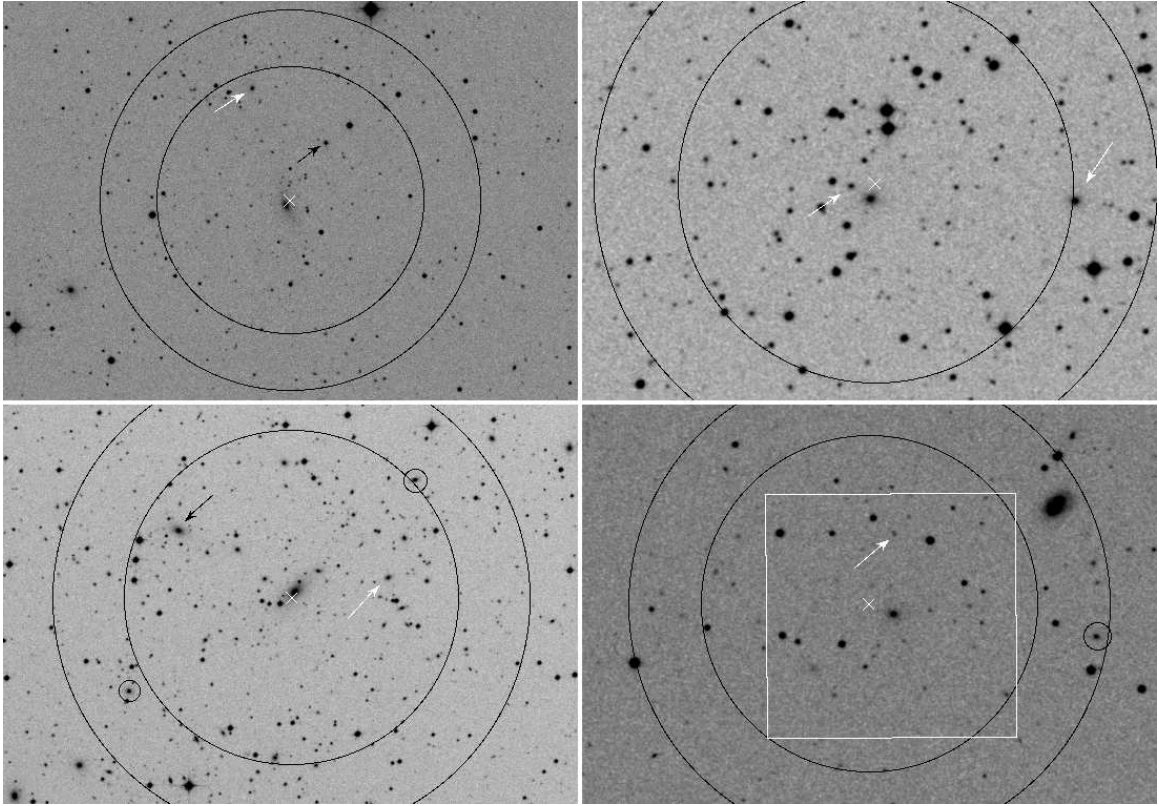}
  \caption{DSS2 images of FGs. From left to right top to bottom:
    cl1416p2315, cl2114m6800, cl2220m5228, and cl2247p0337.}
 \label{fig:fg9101112}
\end{figure}

\subsubsection*{cl1416p2315}

cl1416p2315 is the first fossil cluster to be discovered. It was
studied by different research groups
(\citealt{2003MNRAS.343..627J,2006AJ....132..514C,2006MNRAS.369.1211K}).
Since SDSS data is available for this object, which is at $z=0.138$,
we also used it as a reference for deciding our FG criteria. The
second brightest galaxy is marked by a white arrow in
Figure~\ref{fig:fg9101112} with $\Delta m_{12}=1.7$. The black arrow
points to a galaxy which according to~\cite{2006AJ....132..514C} does
not belong to the cluster.\footnote{Actually, there are two galaxies
  (see SDSS DR6 data, or~\citealt{2006AJ....132..514C}). We are
  referring to the galaxy with coordinates: $14^h16^m21.745^s$
  $+23^\circ17^m20.97^s$. The second galaxy is significantly dimmer.}
Increasing the search radius to $r_{500}$ does not reveal any more
bright galaxies, which might have changed the classification of the
system as an FG (see also the third panel in
Figure~\ref{fig:magdist}).

\subsubsection*{cl2114m6800}

cl2114m6800 was originally identified as an OLEG
~\cite{1999ApJ...520L...1V}.  The system, which is at $z=0.130$,
satisfies our FG criteria as well. The second brightest galaxy is
marked by a white arrow (see Figure~\ref{fig:fg9101112}) with $\Delta
m_{12}=1.99$.  If we increase the search radius to $r_{500}$, we find
another bright galaxy 2MASX~J21134604$-$6801297, which does not obey
$\Delta m_{12}\ge1.7$. Since the redshift of this galaxy is unknown,
it is not clear that it belongs to the group.

\subsubsection*{cl2220m5228}

cl2220m5228 is the seventh new FG discovered.  For this object we do
not have a CCD image, but only a DSS2 image. However, we have found
ENACS data with photometry obtained in the R-filter and measured
redshifts for the central and several other bright galaxies. The group
is at $z=0.102$. In Figure~\ref{fig:fg9101112} the second brightest
galaxy is marked by a white arrow with $\Delta m_{12}=2.21$. The black
arrows point to foreground galaxies. If the search radius would be
increased to $r_{500}$ two more bright galaxies (black circles) would
have to be considered which would make this group a normal group.
Since we do not have redshift information for these galaxies it is
unclear if they in fact belong to the system.

\subsubsection*{cl2247p0337}

cl2247p0337 is another previously discovered
OLEG~\citep{1999ApJ...520L...1V}. It would also pass our criteria for
a fossil system. However, the CCD image from RTT150 does not
completely cover the search region (see the white rectangle in
Figure~\ref{fig:fg9101112}).

The examination of the larger DSS2 image shows that there is no other
bright galaxy inside the search radius.  Therefore, we consider this
object to be an FG (the second brightest galaxy with $\Delta
m_{12}=2.56$ is marked by the white arrow).  If we increase the search
radius up to $r_{500}$, we find a big foreground galaxy NGC~7376. A
visual comparison of angular sizes of this galaxy and the central
cluster galaxy leads to the conclusion that it must be foreground.
Another bright galaxy, which could make this object a normal group, is
marked by a black circle in Figure~\ref{fig:fg9101112}. Since neither
photometry nor redshift of this galaxy are known, we cannot determine
if it belongs to the group or not.


\begin{thebibliography}{99.}

\bibitem[Adelman-McCarthy et al.(2008)]{2008ApJS..175..297A}
  Adelman-McCarthy, J.K., et al.\ 2008, \apjs, 175, 297

\bibitem[Bertin \& Arnouts(1996)]{1996A&AS..117..393B} Bertin, E., \&
  Arnouts, S.\ 1996, \aaps, 117, 393

\bibitem[Blanton \& Roweis(2007)]{2007AJ....133..734B} Blanton, M.R.,
  \& Roweis, S.\ 2007, \aj, 133, 734

\bibitem[Burenin et al.(2007)]{2007ApJS..172..561B} Burenin, R.A.,
  Vikhlinin, A., Hornstrup, A., Ebeling, H., Quintana, H., \&
  Mescheryakov, A.\ 2007, \apjs, 172, 561

\bibitem[Cash(1979)]{1979ApJ...228..939C} Cash, W.\ 1979, \apj, 228,
  939

\bibitem[Cypriano et al.(2006)]{2006AJ....132..514C} Cypriano, E.S.,
  Mendes de Oliveira, C.L., \& Sodr{\'e}, L.J.\ 2006, \aj, 132, 514

\bibitem[Dariush et al.(2007)]{2007MNRAS.382..433D} Dariush, A.,
  Khosroshahi, H.G., Ponman, T.J., Pearce, F., Raychaudhury, S., \&
  Hartley, W.\ 2007, \mnras, 382, 433

\bibitem[Diaz-Gimenez et al.(2008)]{2008arXiv0809.3483D} Diaz-Gimenez,
  E., Ragone-Figueroa, C., Muriel, H., \& Mamon, G.\ 2008,
  arXiv:0809.3483

\bibitem[Fukugita et al.(1996)]{1996AJ....111.1748F} Fukugita, M.,
  Ichikawa, T., Gunn, J.E., Doi, M., Shimasaku, K., \& Schneider,
  D.P.\ 1996, \aj, 111, 1748

\bibitem[Jones et al.(2003)]{2003MNRAS.343..627J} Jones, L.R., Ponman,
  T.J., Horton, A., Babul, A., Ebeling, H., \& Burke, D.~J.\ 2003,
  \mnras, 343, 627

\bibitem[Helsdon \& Ponman(2003)]{2003MNRAS.340..485H} Helsdon, S.F.,
  \& Ponman, T.J.\ 2003, \mnras, 340, 485

\bibitem[Katgert et al.(1998)]{1998A&AS..129..399K} Katgert, P.,
  Mazure, A., den Hartog, R., Adami, C., Biviano, A., \& Perea, J.\
  1998, \aaps, 129, 399

\bibitem[Khosroshahi et al.(2006)]{2006MNRAS.369.1211K} Khosroshahi,
  H.G., Maughan, B.J., Ponman, T.J., \& Jones, L.R.\ 2006, \mnras,
  369, 1211

\bibitem[Khosroshahi et al.(2007)]{2007MNRAS.377..595K} Khosroshahi,
  H.G., Ponman, T.J., \& Jones, L.R.\ 2007, \mnras, 377, 595

\bibitem[Mendes de Oliveira et al.(2006)]{2006AJ....131..158M} Mendes
  de a Oliveira, C.L., Cypriano, E.S., \& Sodr{\'e}, L.J.\ 2006,
  AJ, 131, 158

\bibitem[Mescheryakov et al.(2009), in preparation]{mescheryakov}
  Mescheryakov et al. 2009 (in preparation)

\bibitem[Milosavljevi{\'c} et al.(2006)]{2006ApJ...637L...9M}
  Milosavljevi{\'c}, M., Miller, C.J., Furlanetto, S.R., \& Cooray,
  A.\ 2006, \apjl, 637, L9

\bibitem[Mulchaey \& Zabludoff(1999)]{1999ApJ...514..133M} Mulchaey,
  J.S., \& Zabludoff, A.~I.\ 1999, \apj, 514, 133

\bibitem[Navarro et al.(1997)]{1997ApJ...490..493N} Navarro, J.F.,
  Frenk, C.S., \& White, S.D.M.\ 1997, \apj, 490, 493

\bibitem[La Barbera et al.(2008)]{2008arXiv0812.2929L} La Barbera, F.,
  de Carvalho, R.R., de la Rosa, I.G., Sorrentino, G., Gal, R.R.,
  \& Kohl-Moreira, J.L.\ 2008, arXiv:0812.2929

\bibitem[Osmond \& Ponman(2004)]{2004MNRAS.350.1511O} Osmond,
  J.P.F., \& Ponman, T.J.\ 2004, \mnras, 350, 1511

\bibitem[O'Sullivan et al.(2001)]{2001MNRAS.328..461O} O'Sullivan, E.,
  Forbes, D.A., \& Ponman, T.J.\ 2001, \mnras, 328, 461

\bibitem[Oyaizu et al.(2008)]{2008ApJ...674..768O} Oyaizu, H., Lima,
  M., Cunha, C.E., Lin, H., Frieman, J., \& Sheldon, E.~S.\ 2008,
  \apj, 674, 768

\bibitem[Ponman et al.(1994)]{1994Natur.369..462P} Ponman, T.J.,
  Allan, D.J., Jones, L.R., Merrifield, M., McHardy, I.M., Lehto,
  H.J., \& Luppino, G.A.\ 1994, Nature, 369, 462

\bibitem[Press \& Schechter(1974)]{1974ApJ...187..425P} Press, W.H.,
  \& Schechter, P.\ 1974, \apj, 187, 425

\bibitem[Romer et al.(2000)]{2000ApJS..126..209R} Romer, A.K., et
  al.\ 2000, \apjs, 126, 209

\bibitem[Santos et al.(2007)]{2007AJ....134.1551S} Santos, W.A.,
  Mendes de Oliveira, C., \& Sodr{\'e}, L.J.\ 2007, \aj, 134, 1551

\bibitem[Schechter(1976)]{1976ApJ...203..297S} Schechter, P.\ 1976,
  \apj, 203, 297

\bibitem[Sun et al.(2004)]{2004ApJ...612..805S} Sun, M., Forman, W.,
  Vikhlinin, A., Hornstrup, A., Jones, C., \& Murray, S.~S.\ 2004,
  \apj, 612, 805

\bibitem[Vikhlinin et al.(1999)]{1999ApJ...520L...1V} Vikhlinin, A., 
McNamara, B.R., Hornstrup, A., Quintana, H., Forman, W., Jones, C., 
\& Way, M.\ 1999, \apjl, 520, L1

\bibitem[Vikhlinin et al.(1998)]{1998ApJ...502..558V} Vikhlinin, A.,
  McNamara, B.R., Forman, W., Jones, C., Quintana, H., \& Hornstrup,
  A.\ 1998, \apj, 502, 558

\bibitem[Vikhlinin et al.(2008)]{2008arXiv0805.2207V} Vikhlinin, A., et 
al.\ 2008, arXiv:0805.2207

\bibitem[Voevodkin et al.(2002)]{2002AstL...28..366V} Voevodkin,
  A.A., Vikhlinin, A.A., \& Pavlinsky, M.N.\ 2002, Astronomy
  Letters, 28, 366

\bibitem[Voevodkin et al.(2008)]{2008ApJ...684..204V} Voevodkin, A.,
  Miller, C.J., Borozdin, K., Heitmann, K., Habib, S., Ricker, P., \&
  Nichol, R.C.\ 2008, \apj, 684, 204


\bibitem[von Benda-Beckmann et al.(2008)]{2008MNRAS.386.2345V} von
  Benda-Beckmann, A.M., D'Onghia, E., Gottl{\"o}ber, S., Hoeft, M.,
  Khalatyan, A., Klypin, A., M{\"u}ller, V.\ 2008, \mnras, 386,
  2345

\bibitem[Yoshioka et al.(2004)]{2004AdSpR..34.2525Y} Yoshioka, T.,
  Furuzawa, A., Takahashi, S., Tawara, Y., Sato, S., Yamashita, K., \&
  Kumai, Y.\ 2004, Advances in Space Research, 34, 2525

\bibitem[Zibetti et al.(2008)]{2008arXiv0809.2036Z} Zibetti, S.,
  Pierini, D., \& Pratt, G.W.\ 2008, arXiv:0809.2036

\end{thebibliography}
\end{document}